\documentclass[12pt]{mypaper}
\pdfoutput=1
\usepackage{graphicx}
\usepackage{rotate}
\usepackage{rotating}
\usepackage{relsize}
\usepackage{slashed}
\usepackage{url}
\usepackage{amsmath}
\usepackage{longtable}
\usepackage{setspace} 
\usepackage[font=small]{caption}
\usepackage[bottom]{footmisc}
    
\usepackage{latexsym}
\usepackage{amssymb}
\usepackage{epsfig,amsmath,graphics}
\usepackage{color}
\usepackage{dcolumn}
\usepackage{slashed}
\usepackage{comment,latexsym} 
\usepackage{bm}
\usepackage{verbatim}
\usepackage{tabularx}
\usepackage{dcolumn}
\usepackage{hyperref}
\usepackage{setspace}
\usepackage{xspace}
\usepackage{bigints}
\usepackage{multirow}
\usepackage{cite}

\usepackage{cancel}

\def\be{\begin{equation}}
\def\ee{\end{equation}}
\newcommand{\beq}{\begin{equation}}
\newcommand{\eeq}{\end{equation}}
\def\bea{\begin{eqnarray}}
\def\eea{\end{eqnarray}}

\newcommand{\GeV}{{\text{ GeV}}}
\newcommand{\TeV}{{\text{ TeV}}}

\newcommand{\ZTwo}{\ensuremath{\mathbf{Z}_2\ }}

\begin{document}


%
\begin{titlepage}
\flushright{RU-NHETC-2013-26}
\title{{\Large Doubling down on naturalness with \\ a supersymmetric twin Higgs}}
\vspace{-30pt}
 \begin{Authlist}
Nathaniel Craig
\vspace{-5pt}
\Instfoot{rutgers}{New High Energy Theory Center \\ Department of Physics and Astronomy \\Rutgers University \\ Piscataway, NJ 08854}
Kiel Howe
\vspace{-5pt}
\Instfoot{stanford}{Stanford Institute for Theoretical Physics \\ Department of Physics \\ Stanford University \\ Stanford, CA 94305}
\Instfoot{SLAC}{SLAC National Accelerator Laboratory \\ Menlo Park, CA  94025 USA}
\end{Authlist}

\begin{abstract}

We show that naturalness of the weak scale can be comfortably reconciled with both LHC null results and observed Higgs properties provided the double protection of supersymmetry and the twin Higgs mechanism. This double protection radically alters conventional signs of naturalness at the LHC while respecting gauge coupling unification and precision electroweak limits. We find the measured Higgs mass, couplings, and percent-level naturalness of the weak scale are  compatible with stops at $\sim 3.5$ TeV and higgsinos at $\sim 1$ TeV. The primary signs of naturalness in this scenario include modifications of Higgs couplings, a modest invisible Higgs width, resonant Higgs pair production, and an invisibly-decaying heavy Higgs.

\end{abstract}
\end{titlepage}

\setstretch{1.1}
\section{Introduction}
Just as the discovery of a Standard Model (SM)-like Higgs boson at the LHC \cite{Aad:2012tfa, Chatrchyan:2012ufa} sharpens the urgency of the hierarchy problem, the onward march of null results in searches for new physics places increasing stress upon conventional ideas for electroweak naturalness. Perhaps electroweak naturalness is a dead end, with the solution to the hierarchy problem lying somewhere in the landscape. But perhaps electroweak naturalness is still close at hand, concealed only by its unexpected properties. This latter possibility raises a pressing question: {\it Can we learn anything new about electroweak naturalness from null results at the LHC?} More specifically,
\begin{itemize}
\item Are there signatures of naturalness other than conventional top partners? 
\item Are there wholly natural theories where the conventional signs of naturalness -- especially supersymmetric naturalness --  may be out of reach of the LHC? 
\end{itemize}

To a certain extent these possibilities are illustrated by composite Higgs models \cite{Kaplan:1983fs, Kaplan:1983sm}, where the Higgs mass is protected by a global symmetry and heavy resonances lie in the multi-TeV range  -- but even here, one expects light fermionic top partners to accommodate the observed Higgs mass \cite{Matsedonskyi:2012ym}, as well as copious production of heavy resonances in the second LHC run. Moreover, such models are typically in tension with precision electroweak constraints and hints of gauge coupling unification, at variance with what few indirect indications we have about physics in the ultraviolet. While there is also still room for conventional supersymmetric models with light superpartners whose signatures are muddled by reduced event activity \cite{LeCompte:2011cn} or missing energy \cite{Fan:2011yu}, these solutions are under increasing pressure from evolving search strategies at the LHC -- and, in any event, have little intrinsic connection between naturalness and the lack of natural signals.

An attractive possibility is to consider theories enjoying {\it double protection} of the Higgs potential, for example via both supersymmetry and a spontaneously broken global symmetry \cite{Birkedal:2004xi, Chankowski:2004mq}. This raises the prospect of partially decoupling the signals of each symmetry mechanism without imperiling the naturalness of the weak scale. In this work we explore the double protection provided by the combination of the twin Higgs mechanism \cite{Chacko:2005pe} and supersymmetry.\footnote{For recent work in a similar spirit combining supersymmetry and the composite Higgs, see e.g. \cite{Caracciolo:2012je}.} In these models an exact \ZTwo symmetry between the MSSM and a mirror MSSM leads to an approximate U(4) symmetry, and the light Higgs is primarily composed of the pseudo-goldstones of the broken U(4).  Although supersymmetry plays a role in the ultraviolet completion, the stops need not be light. Moreover, the fermionic top partner furnished by double protection is {\it neutral under the Standard Model gauge group}. Rather, the predominant signals of naturalness emerge through the Higgs portal: modifications of Higgs couplings, an invisible Higgs width, resonant Higgs pair production, and an invisibly-decaying heavy Higgs. Thanks to double protection of the Higgs potential, the conventional signs of supersymmetric naturalness are absent even at the 13/14 TeV LHC, with percent-level tuning in the Higgs vev (comparable to the ``fine-tuning'' of the QCD scale) compatible with stops at $\sim 3.5$ TeV and higgsinos at $\sim 1$ TeV.

The supersymmetric UV completion of the twin Higgs model we study has the attractive features of maintaining perturbative gauge coupling unification, calculable and safe precision electro-weak and flavor observables, and a light CP-even Higgs mass naturally in the experimentally observed window. While many interesting conclusions about twin Higgs phenomenology can be reached from an effective theory of only the scalar Higgs and SM degrees of freedom in twin models\cite{Barbieri:2005ri}, studying a full UV completion also has the advantage of an unambiguous tuning measure to compare to other perturbative solutions to the naturalness problem like the NMSSM and a direct understanding of collider limits on all of the new colored and electroweak states. Although supersymmetric completions of mirror and left-right twin Higgs models were considered prior to Higgs discovery \cite{Chang:2006ra,Falkowski:2006qq}, they focused on eliminating the intrinsic tuning from supersymmetric quartics at the cost of additional model-building complexity and a loss of MSSM-like gauge coupling unification. In this work we explore the simplest supersymmetric mirrow twin Higgs in light of the observed mass and couplings of the SM-like Higgs, taking the tuning arising from supersymmetric quartics at face value. Venturing beyond the pseudo-goldstone limit and accounting for the contributions of the full Higgs effective potential, we find that this simple model has tuning comparable to the more complicated efforts.

Our paper is organized as follows: In section \ref{sec:model} we begin by reviewing the simplest supersymmetric twin Higgs model and the parametrics of the Higgs potential in the pseudo-goldstone limit. We then turn to an analysis of the Higgs mass and full effective potential at one loop, computing the fine-tuning of the theory as a function of the superpartner mass scales.  In section \ref{sec:pheno} we study the phenomenology of the model in light of the Higgs discovery, focusing on the implications of Higgs couplings for the allowed parameter space and detailing the most relevant signals of naturalness. In section \ref{sec:limits} we consider ancillary limits from precision electroweak, flavor, and cosmological considerations. We reserve a detailed discussion of the possible UV completions of the twin Higgs singlet portal for the appendix.

\section{A Supersymmetric Twin Higgs}
\label{sec:model}

\subsection{Basic Set-up}
\label{sec:setup}
Mirror Twin Higgs models are based on the idea that a \ZTwo symmetry exchanging the SM Higgs and a ``mirror" Higgs field charged under a distinct identical copy of the Standard Model gauge group leads to an accidental U(4) symmetry in the quadratic terms of the Higgs potential \cite{Chacko:2005pe,Barbieri:2005ri}. If the \ZTwo symmetry is exact -- implying a complete mirror copy of the matter and gauge fields coupled to the mirror Higgs -- then the full quadratic effective potential including UV-sensitive mass corrections possesses the accidental U(4) symmetry. The light SM Higgs doublet is identified with some of the pseudo-goldstones of the spontaneously broken U(4) and is therefore protected from quadratic sensitivity to the cutoff. The sensitivity to UV scales only re-emerges through the (presumably small) quartic and higher order terms explicitly breaking the U(4) symmetry, and is suppressed by the (presumably large) coefficient of the U(4) preserving quartic terms in a perturbative completion (or equivalently $\sim(4\pi)^2$ in a composite model). As with any pseudo-goldstone mechanism for protecting the Higgs mass, a UV completion such as supersymmetry is required for the theory above a few TeV.

Our perturbative SUSY twin Higgs model comprises two complete copies of the $\rm MSSM$, an ``A-sector" which will correspond to the observed sector with the light fields of the Standard Model, and a ``B-sector" with identical copies of the MSSM gauge group and field content. The couplings and soft SUSY breaking masses of the two sectors are set equal by a \ZTwo symmetry exchanging the A and B sectors. A single singlet superfield $S$ couples the A and B sectors. The combination of supersymmetry and the \ZTwo symmetry yields a theory that is, in principle, complete up to the Planck scale.

The \ZTwo and gauge symmetries guarantee that the singlet-Higgs interactions respect a full U(4) global symmetry, of which gauge and Yukawa interactions preserve an ${\rm SU(2)_A \times U(1)_A \times SU(2)_B \times U(1)_B}$ subgroup.  To make this explicit, we write the A and B sector Higgs fields  in U(4) multiplets as
\begin{equation}
H_u = \begin{pmatrix} h^A_u \\ h^B_u \end{pmatrix}, \;\;\; H_d = \begin{pmatrix} h^A_d \\ h^B_d \end{pmatrix},
\end{equation}
and the superpotential of the Higgs-singlet sector becomes
\begin{eqnarray}
W_{\rm U(4)} &=& \mu (h^A_u h^A_d  + h^B_u h^B_d) + \lambda S (h^A_u h^A_d  + h^B_u h^B_d) + M_S S S  \nonumber \\
		     &\equiv& \mu H_u H_d + \lambda S H_u H_d + M_S S S. 
\end{eqnarray}

The \ZTwo symmetry also guarantees that the quadratic soft breaking terms preserve the full U(4), even after radiative corrections. Assuming the singlet has a large soft mass $m^2_S \gg \mu, M_S$, it will decouple leaving its F-term quartic intact. The full U(4) preserving scalar potential in the Higgs sector is then given by the sum of supersymmetric and soft contributions,
\begin{equation}
V_{\rm U(4)} = (m^2_{H_u}+\mu^2) |H_u|^2 + (m^2_{H_d} + \mu^2) |H_d|^2 - b ( H_u H_d + {\rm h.c.}) + \lambda^2 |H_u H_d|^2 \label{eq:VU4}
\end{equation}
Crucially, the \ZTwo symmetry automatically leads to a potential for the Higgs with both quadratic terms  \emph{and a potentially large quartic term} respecting the larger U(4) symmetry. This can be contrasted with composite twin Higgs models, where the \ZTwo  on its own does not guarantee that the strong sector will respect the necessary U(4) symmetry.

The gauge and Yukawa couplings of the A and B sectors give rise to explicit breaking of the U(4) at both tree and loop level.  When the U(4)-symmetric quartic dominates over the U(4)-breaking quartic terms (and other higher order terms), this model provides a perturbative realization of the twin Higgs mechanism. In particular, in the limit that the $H_u$ and $H_d$ vevs lie completely in the B-sector direction, the pseudo-goldstones of the broken U(4) correspond to a light A-sector Higgs doublet with a scalar mass protected by the twin mechanism against large radiative corrections from the top and gauge sectors. 

In the absence of supersymmetry, electroweak gauge and Yukawa interactions would only give rise to U(4) breaking quartics at one loop, the most important of which is the quartic $\delta \lambda_u$ generated by the top sector. However, with the introduction of supersymmetry, the D-terms of the A and B sector gauge groups necessarily generate U(4)-breaking quartic terms at tree level.  For the neutral components of the Higgs field these contributions are
\begin{equation}
V_{\rm \cancel{U(4)}} = \frac{g^2 + g'^2}{8}\left[(|{h^0_u}^A|^2 - |{h^0_d}^A|^2)^2 + (|{h^0_u}^B|^2 - |{h^0_d}^B|^2)^2  \right] + \delta \lambda_u (|{h^0_u}^A|^4 + |{h^0_u}^B|^4)  + \ldots.
\label{eq:VU4Breaking}
\end{equation}

The U(4)-breaking terms are important to generate a mass for the light pseudo-goldstone Higgs, but unfortunately their form necessarily leads to symmetric vevs between the A and B sector, $v_A = v_B$, which we find to be phenomenologically unviable. To rectify this problem, we assume there is a small source of soft breaking of the \ZTwo symmetry which we take to be of the simple form
\begin{equation}
V_{\cancel \ZTwo} = \Delta m^2_{H_u} (|h_u^A|^2 - |h_u^B|^2) + \Delta m^2_{H_d}  (|h_d^A|^2 - |h_d^B|^2).
\label{eq:VZ2Breaking}
\end{equation}

\subsection{The pseudo-goldstone limit}

The Higgs sector of the SUSY twin model can be most easily understood in the limit that all of the non-goldstone directions have decoupled. This condition is satisfied at tree-level when $b \sin2\beta \gg (\frac{g^2 + g'^2}{4} \cos^2 2\beta) f^2$ and $\lambda^2   \gg  \frac{g^2 + g'^2}{2} \cot^2 2\beta$, where $f^2 \equiv \langle h_A \rangle^2 + \langle h_B \rangle^2 \equiv v_A^2 + v_B^2$ is the total magnitude of the U(4) breaking vev.

In this limit, $f$ and $\tan\beta_A = \tan\beta_B$ can be determined from the U(4) symmetric potential (Eq. \ref{eq:VU4}),
\begin{eqnarray}
\tan\beta & = & \frac{\mu^2 + m^2_{H_d}}{\mu^2 + m^2_{H_u}} \\
f^2 & = & \frac{1}{{\lambda^2}}\left(m^2_A - 2\mu^2 - m^2_{H_u} - m^2_{H_d}\right)
\end{eqnarray}
where $m_A=\frac{2b}{\sin2\beta}$ is the tree-level mass of one of the physical pseudoscalar Higgses.
To study the light Higgs state, it is convenient to work in terms of the nonlinear realization of the uneaten goldstone direction
\begin{equation}
H_u = f \sin\beta\begin{pmatrix}  0 \\ \sin\frac{\phi}{\sqrt{2}f} \\  0 \\ \cos\frac{\phi}{\sqrt{2}f}\end{pmatrix}, \;\;\; H_d =  f \cos\beta\begin{pmatrix}  0 \\ \sin\frac{\phi}{\sqrt{2}f} \\  0 \\ \cos\frac{\phi}{\sqrt{2}f}\end{pmatrix}
\end{equation}
where $\phi$ is the pseudo-goldstone Higgs. A potential for $\phi$ is generated by the U(4) breaking terms Eqs.~\ref{eq:VU4Breaking} and \ref{eq:VZ2Breaking}. The minimization conditions yield the vev in terms of the \ZTwo breaking masses
\begin{equation}
\sin^2\frac{\phi}{\sqrt{2}f} = \frac{v^2}{f^2} = \frac{1}{2}\left(1 - \frac{\Delta m^2}{\left(\frac{g^2+g'^2}{8}\cos^2 2\beta + \delta\lambda_u \sin^2\beta\right)f^2}\right)
\label{eq:goldstoneminimum}
\end{equation}
where we now take the canonical observed vev in the A sector $v_A = v \approx 174 \GeV$ and define $\Delta m^2 \equiv \Delta m^2_{H_u}  \sin^2\beta + \Delta m^2_{H_d}  \cos^2\beta$. The mass of the light state $\phi$ at the minimum is given by
\begin{equation}
m^2_\phi = (m_Z^2 \cos^2 2\beta + 4\delta\lambda_u v^2 \sin^4\beta)\left(2 - \frac{2v^2}{f^2}\right).
\label{eq:goldstonemass}
\end{equation}

Eqs.~\ref{eq:goldstoneminimum} and \ref{eq:goldstonemass} illustrate several important points for the following more detailed discussion. First, it is clear that to obtain a hierarchy in vevs $v^2 < f^2/2$, the \ZTwo breaking mass terms must be tuned against the potential generated by the U(4) breaking quartic terms. This leads to an intrinsic tuning of the weak scale of order $f^2/2v^2$. Ref. \cite{Chang:2006ra} sought to remedy this tuning in a similar SUSY twin model by removing the B-sector D-term quartics. This additional \ZTwo breaking modifies Eq.~\ref{eq:goldstoneminimum} to give a small hierarchy $v^2 < f^2/2$ even in the absence of a \ZTwo beaking mass, but we see immediately that the remaining symmetric radiative contributions $\delta\lambda_u$ will remain important, and we find numerically that there is in fact very little to be gained by this modification. Likewise ref. \cite{Falkowski:2006qq} sought in a left-right twin SUSY model to introduce a natural hierarchy $v^2 < f^2/2$ through removing the D-term contributions by forcing $\tan\beta = 1$ and including soft \ZTwo breaking quartics from a non-minimal singlet sector. This mechanism can be adapted to the mirror model, but again we find that after including the radiatively generated quartic terms there is little benefit.  In this respect, the added model-building complications of \cite{Chang:2006ra, Falkowski:2006qq} can be sidestepped without substantially worsening the tuning of the theory.

Another important point is that the mass of the light Higgs state is generated by the same quartic terms that give mass to the light MSSM Higgs, with no contributions from the U(4) symmetric coupling $\lambda$. However, for large hierarchies of $v^2/f^2$ there can be up to a factor of two enhancement in the squared mass compared to the MSSM formula, as is evident in Eq.~\ref{eq:goldstonemass}. Physically, in this limit the $\phi$ potential receives contributions from both the A- and B-sector quartics. This enhancement brings the tree-level Higgs mass prediction tantalizingly close to the observed value, and is critical to obtain the observed mass of the SM-like Higgs in regions of small $\tan\beta$. Note also that the MSSM-like limit {\it cannot} be obtained simply by taking the $f \to \infty$ limit of Eq.~\ref{eq:goldstonemass}, since there are large trilinear couplings of $\mathcal{O}(f)$ in the Higgs sector. The MSSM-like limit is instead obtained by taking $\lambda \to 0$ and $M_S \to \infty$, which introduces appropriate corrections to Eq.~\ref{eq:goldstonemass} that are not apparent in the pseudo-goldstone limit.

\subsection{Full effective potential and Higgs mass}
\label{sec:effpot}

Perturbativity limits the range of allowed singlet couplings $\lambda$, and the observed light Higgs mass $m_h\approx125 \GeV$ limits the range of allowed $\tan\beta$. We therefore find that over most of the parameter space of interest there are important non-decoupling effects in the potential and a treatment beyond the pseudo-goldstone limit is necessary.

The structure of the radiative corrections is also very important to understanding the light Higgs mass and the minimum of the U(4) breaking potential, and we find it is necessary to carefully include the large U(4) breaking contributions to the effective potential. In particular, we evaluate the effective potential at the SUSY breaking scale $m_{\rm soft}$ including the full leading log plus one-loop finite contributions from both the A and B top/stop sectors (see e.g. \cite{Carena:1995bx, Carena:1995wu}), as well as the one-loop leading log contributions from the A and B electroweak gauge sectors. The leading contributions of the singlet to the effective potential are U(4) symmetric and not included in our analysis. A qualitatively important aspect of the effective potential is that it is \ZTwo symmetric and has a minimum at the symmetric vev $v_A = v_B$, as can easily be seen from inspecting the one-loop contributions. Therefore the \ZTwo breaking masses remain necessary to obtain a hierarchy in vevs.

After including the effective potential contributions to the full tree-level potential of Eqs.~\ref{eq:VU4}, \ref{eq:VU4Breaking}, and \ref{eq:VZ2Breaking}, we numerically determine the minimum and spectrum of Higgs states, including the wave-function renormalization of the lightest Higgs state. Away from the pseudo-goldstone limit $\tan\beta_A=\tan\beta_B$ no longer necessarily holds. We fix the relative values of the \ZTwo breaking masses by requiring $\tan\beta_B = \tan\beta_A - 0.1$, which leads typically to a similar magnitude for $\Delta m^2_{H_u}$ and $\Delta m^2_{H_d}$.

\begin{figure}[t]
  \centering
  \includegraphics[width=.48\columnwidth]{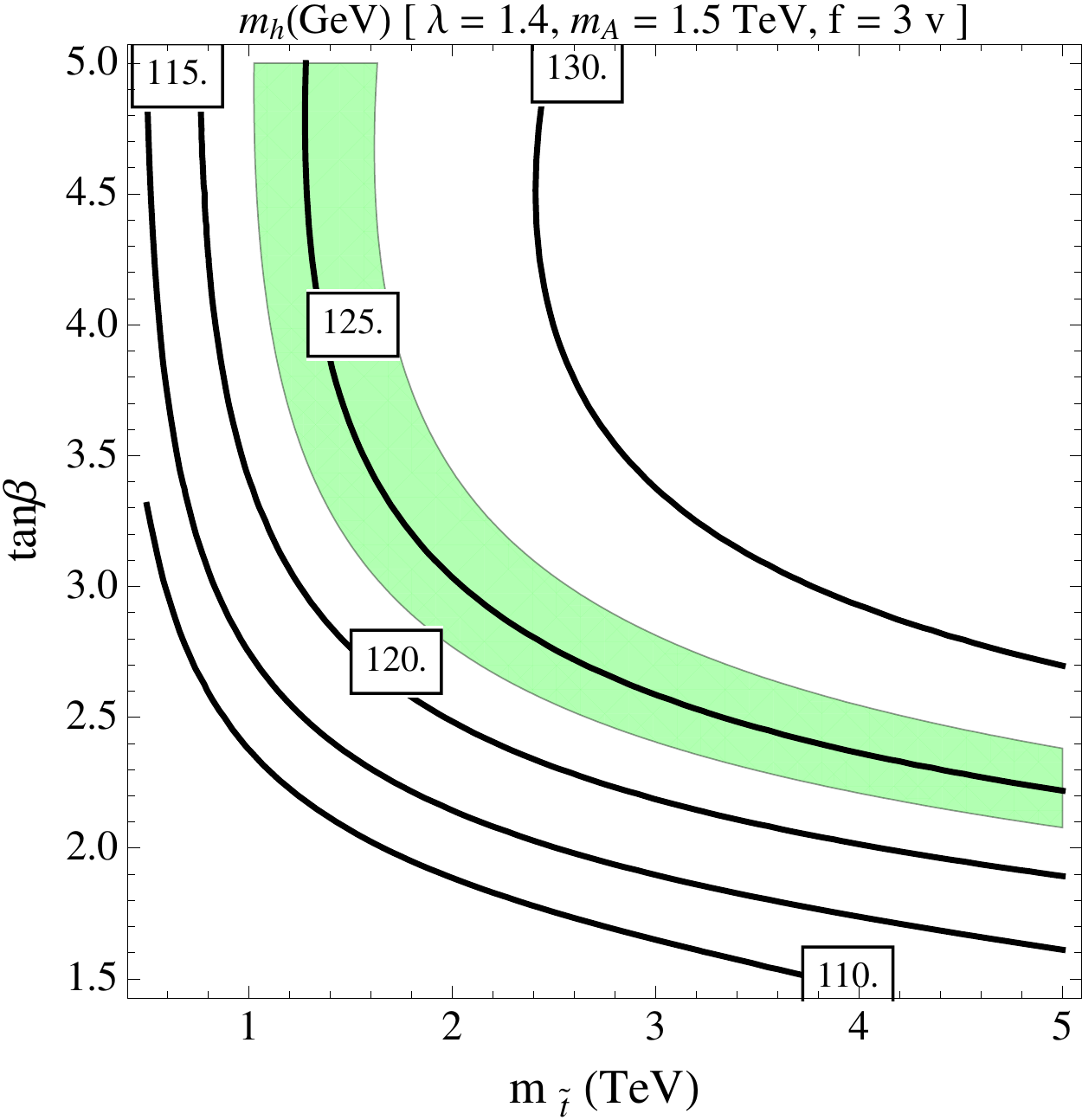}
  \caption{The lightest Higgs mass in the SUSY twin Higgs model as a function of a common stop mass $m_{\tilde t_1} = m_{\tilde t_2} \equiv m_{\tilde t}$ and $\tan \beta$ with $\lambda = 1.4$, $f=3v$, and $m_A = 1.5 \TeV$. The green shaded region denotes $123 \GeV < m_{h} < 127 \GeV$. }
    \label{fig:HiggsMass}
\end{figure}

The light Higgs mass for  $f=3v$, $\lambda=1.4$, and $m_A = 1.5 \TeV$ is shown in Fig.~\ref{fig:HiggsMass} as a function of $\tan\beta$ and a common stop mass $m_{\tilde t_1} = m_{\tilde t_2}$ with no mixing. We find that the non-decoupling effects decrease the mass by $5-10\%$ below the pseudo-goldstone expectation of Eq.~\ref{eq:goldstonemass} in the region of interest. For large $\tan\beta$, the radiative corrections from $m_{\tilde t}\approx 1.3\TeV$ stops are necessary to obtain the observed Higgs mass, while for a heavy stop $m_{\tilde t}\approx 5\TeV$, $\tan\beta$ can be as small as $2.4$.

\subsection{Fine-tuning}

The supersymmetric UV completion of the mirror twin Higgs model provides the crucial advantage of allowing a meaningful calculation of fine-tuning in terms of soft SUSY-breaking parameters.  There are two independent sources of tuning in the twin SUSY model. The first comes from creating a hierarchy between the A and B sector vevs. According to Eq.~\ref{eq:goldstoneminimum}, this introduces a tuning of $\Delta m^2$ against the quartic U(4) breaking terms,
\begin{equation}
\label{eq:deltavfApprox}
	\Delta_{v/f} \approx \frac{\partial \ln (v^2/f^2)}{\partial  \Delta m^2} = \left(\frac{f^2}{2v^2}-1\right).
\end{equation}
Numerically we find this relationship to be quite accurate even away from the pseudo-goldstone limit and in the presence of additional contributions to the effective potential. This tuning is present in any twin Higgs model in which a soft \ZTwo breaking mass leads to the hierarchy in vevs \cite{Barbieri:2005ri}. An important aspect of the twin mechanism is that the \ZTwo breaking is soft and therefore the $\Delta m^2$ terms do not have any additive sensitivity to other soft masses.

The second source of tuning in the twin SUSY model is the tuning of the total U(4) breaking vev $f$ against the quadratic contributions to the U(4) symmetric Higgs masses. This is analogous to the tuning of the normal electroweak vev in non-twinned SUSY models.  At one loop the most important radiative corrections to the U(4) symmetric Higgs masses arise from the  stop and singlet soft masses
\begin{eqnarray}
\delta m^2_{H_u} &\approx& \frac{3y_t^2}{8\pi} (m^2_{\tilde t_L}+ m^2_{\tilde t_R}) \log \frac{\Lambda_{\rm mess}}{m_{\rm soft}} + \frac{\lambda^2}{8\pi^2}m^2_S \log \frac{\Lambda_{\rm mess}}{m_{\rm soft}}  + \dots \\ 
\delta m^2_{H_d} &\approx& \frac{\lambda^2}{8\pi^2}m^2_S \log \frac{\Lambda_{\rm mess}}{m_{\rm soft}}  + \dots
\end{eqnarray}
where $\Lambda_{\rm mess}$ is the scale of mediation of SUSY breaking. In the full RG there are also important contributions from the effect of the gluino on the running of the stop mass and from the running of $\lambda$ if it approaches its Landau pole near the messenger scale. 

In the limit $m^2_A \gg \lambda^2 f^2$, the $f$ tuning takes a simple form. When the dominant tuning is due to the stop contributions to the up-type Higgs mass for example,
\begin{equation}
\label{eq:deltafApprox}
\Delta_f \approx \frac{\partial \ln f^2}{\partial \ln \delta m^2_{H_u}} = \frac{\delta m^2_{H_u}}{2 \lambda^2 f^2 \cos^2 \beta}.
\end{equation}

The total tuning of the the SUSY twin model is the product of the two independent tunings:
\begin{equation}
\label{eq:TwinTuningApprox}
\Delta_{\rm twin} = \Delta_f  \times \Delta_{v/f} \underset{v^2\ll f^2}{\approx} \frac{\delta m^2_{H_u}}{4\lambda^2 v^2 \cos^2\beta} 
\end{equation}
where we have taken the approximate expressions Eqs.~\ref{eq:deltavfApprox} and \ref{eq:deltafApprox} in the limit $v^2\ll f^2$.

It is interesting to measure the relative improvement in tuning of the SUSY twin model compared to a more minimal alternative. A convenient benchmark is the NMSSM, which can likewise accommodate the observed Higgs mass with tree-level contributions from the singlet quartic and has been shown to compare favorably with a number of alternative models for reducing the tuning of SUSY models in light of recent LHC results \cite{Arvanitaki:2013yja}. The NMSSM tuning equation has nearly identical form to the tuning of the SUSY twin model in the same decoupling limit,
\begin{equation}
\label{eq:NMSSMTuningApprox}
\Delta_{\rm NMSSM} \approx \frac{\delta m^2_{H_u}}{2 \lambda_{\rm NMSSM}^2 v^2 \cos^2\beta} \approx \frac{\delta m^2_{H_u}}{m^2_{h} / (2\sin^2\beta)}.
\end{equation}
The key difference is that in the NMSSM, the value of the quartic coupling in the denominator is fixed by the observed light Higgs mass, $m_h\approx 125\GeV$. In the SUSY twin model, the twin mechanism protects the light A-sector pseudo-goldstone Higgs mass from the large U(4) invariant quartic coupling $\lambda$ in the denominator. The tuning can therefore be substantially reduced while maintaining a light Higgs. For example, for $\tan\beta = 2$ and $\lambda = 1.4$, Eqs.~\ref{eq:TwinTuningApprox} and \ref{eq:NMSSMTuningApprox} imply that the NMSSM is roughly five times more tuned than the SUSY twin model for the same stop mass. Similar relationships holds for the relative tuning with respect to the singlet soft mass and the tree level $\mu$-term.

This discussion also brings up an important difference between the SUSY twin model and composite twin Higgs models. In composite twin Higgs models, the connection between the tuning and the light Higgs mass re-enters only through the logarithmic dependence on the cut-off scale for the radiative U(4) breaking quartic terms. On the other hand in the SUSY twin model, the simultaneous requirement of a perturbative singlet coupling $\lambda$ and the observed light Higgs mass introduces an indirect constraint on the size of the effective quartic coupling setting the $f$ tuning. In detail, the structure of the D-term and radiative contributions to the U(4) breaking quartic terms fixes $\tan \beta$ for a given Higgs mass and set of soft parameters. However, the effective size of the tree-level U(4) preserving quartic coupling is dependent on $\tan\beta$ and enters the tuning formulae in the decoupling limit as $\lambda^2 \left(\frac{\sin2\beta}{2}\right)^2$. For perturbative couplings $\lambda \lesssim 2$, it's critical that the correct Higgs mass can arise at small values of $\tan\beta$ to obtain a large effective quartic.


To improve upon these rough estimates of tuning, we perform a numerical study of the parameter space using the full one-loop RG equations and the complete Higgs effective potential as described in Sec.~\ref{sec:effpot}. In particular, we define a point in the low energy parameter space with a choice of the parameters $\lambda$, $f$, $m^2_{\tilde{t}}$, $m^2_{\tilde{S}}$, $m_{A}$, $\mu$, and $\tan \beta$ defined at the scale $m^2_{\rm soft} =m^2_{\tilde{t}}$.  For simplicity, at the soft scale  we take the limit of no stop mixing and degenerate stop masses $m_{\tilde{t_L}} =m_{\tilde{t_R}}\equiv m_{\tilde{t}} $ and set the gluino degenerate with the stops, $M_{3} = m_{\tilde{t}}$. We then determine the \ZTwo preserving Higgs soft masses $m^2_{H_u}$ and $m^2_{H_d}$ and the \ZTwo breaking soft masses $\Delta m^2_{H_d}$ and $\Delta m^2_{H_d}$ by minimizing the effective potential.  The tunings of the $f$ and $\frac{v}{f}$ parameters are evaluated by independently varying the soft masses at $\Lambda_{\rm mess}$, running them back down to the soft scale, and numerically evaluating the shift in the vevs,
\begin{eqnarray}
\Delta_f & = & \left[\sum_{x=\{m^2_{\tilde{t_L}}, m^2_{\tilde{t_R}}, M_3, m^2_{H_u}, m^2_{H_d}, \mu, m^2_{\tilde{S}}\}} \left(\frac{\partial \ln f^2|_{m_{\rm soft}}}{\partial \ln x|_{\Lambda_{\rm mess}}}\right)^2\right]^{\frac{1}{2}} \\
\Delta_{\frac{v}{f}} & = & \left[\sum_{x=\{\Delta m^2_{H_u}, \Delta m^2_{H_d}\}} \left(\frac{\partial \ln v^2/f^2|_{m_{\rm soft}}}{\partial \ln x|_{\Lambda_{\rm mess}}}\right)^2\right]^{\frac{1}{2}}
\end{eqnarray}
Note that the twin mechanism protects the running of the \ZTwo breaking masses from additive contributions from the stop and singlet sectors above the soft scale, and the running of the  \ZTwo breaking masses is a small effect on the tuning. Again the combined tuning of the twin model is the product $\Delta_{\rm twin} = \Delta_f  \times \Delta_{\frac{v}{f}}$.

 For comparison we also define a benchmark NMSSM model with the same field content, superpotential, and soft terms as the A-sector plus singlet of the twin model. We use the same framework to determine the low energy parameters of this model and to calculate the tuning, which we define as
\begin{equation}
\Delta^{\rm NMSSM}  =  \left[\sum_{x=\{m^2_{\tilde{t_L}}, m^2_{\tilde{t_R}}, M_3, m^2_{H_u}, m^2_{H_d}, \mu, m^2_{\tilde{S}}\}} \left(\frac{\partial \ln v^2|_{m_{\rm soft}}}{\partial \ln x|_{\Lambda_{\rm mess}}}\right)^2\right]^{\frac{1}{2}}. 
\end{equation} 

In all plots we choose a reference value of the messenger scale  of $\Lambda_{\rm mess} = 100 m_{\tilde t}$, so that for each choice of $m_{\tilde t}$ and $\lambda$ at the soft scale, the value of $\lambda$ at the messenger scale is roughly the same.


\begin{figure}[h!]
  \centering
  \includegraphics[width=.48\columnwidth]{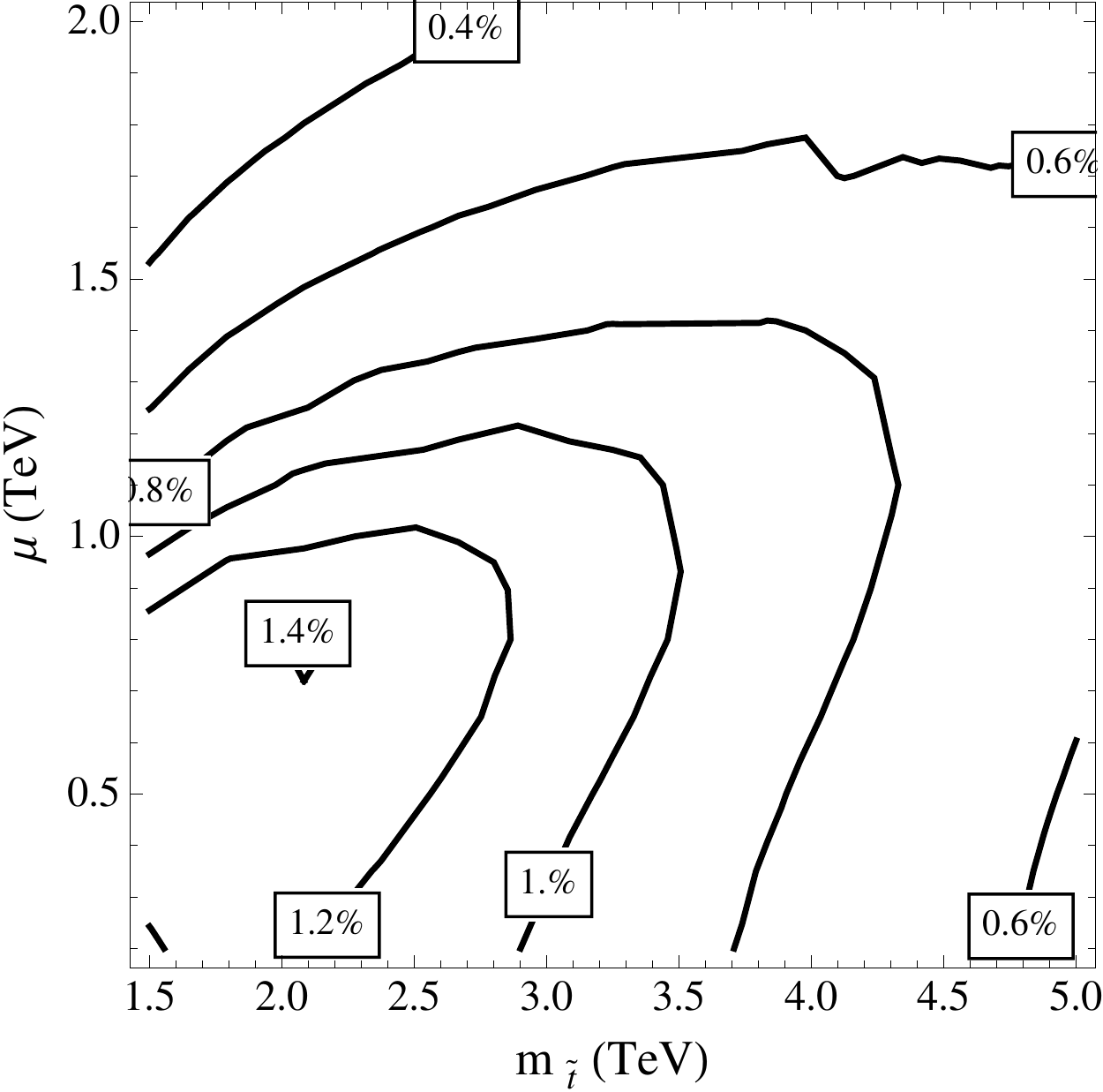}
  \includegraphics[width=.48\columnwidth]{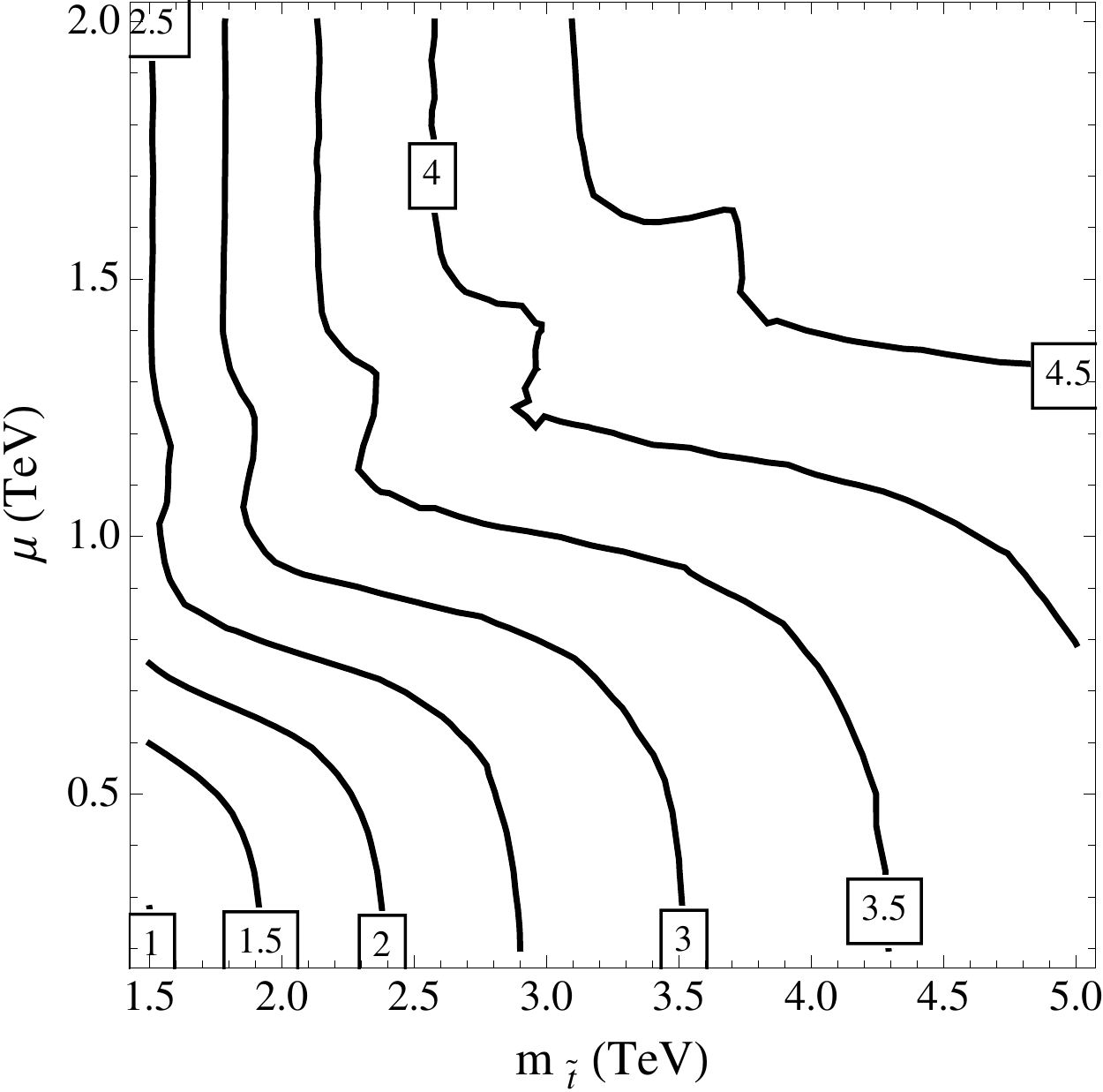}
  \caption{Tuning in the twin SUSY model with $\lambda = 1.4$, $f=3v$, $m_A = 1.5 \TeV$, and $m^2_S = (1 \TeV)^2$. The left is the absolute tuning, and the right is the relative tuning compared to the NMSSM, $\Delta^{\rm NMSSM}/\Delta^{\rm twin}$, with the NMSSM parameters $\lambda = 0.6$, $m_A = 0.8 \TeV$, and $m^2_S = (1 \TeV)^2$. At each point, $\tan\beta$ is determined independently for the twin and NMSSM models to obtain $m_h = 125 \GeV$.}
    \label{fig:tuningStopsHiggsinos}
\end{figure}

The dominant LHC limits on SUSY models come from constraints on the production of colored particles. The mass of the (N)LSP is also important both for direct searches and to determine the sensitivity to the decays of colored particles\footnote{For low-scale mediation models with a light gravitino LSP, the effect of the NLSP mass on limits for colored particles is much less decisive.}. Direct constraints on stops as well as constraints on other colored sparticles\footnote{In the simplest models of SUSY breaking the mass scale for the other colored sparticles must be similar to the stop mass, and searches for these particles set the most stringent constraints. In general at least the gluino mass must be within a factor of $\sim 2$ of the lightest stop mass to avoid introducing additional fine-tuning to obtain a separation after RG flow \cite{Arvanitaki:2012ps}.} therefore enter the tuning through $m_{\tilde{t}}$, while limits on the LSP mass enter the fine-tuning through the tree-level contributions from $\mu$, which must be at least as large as the (N)LSP mass. In Fig.~\ref{fig:tuningStopsHiggsinos} we study the tuning of the twin SUSY model as a function of $\mu$ and $m_{\tilde{t}}$, both in absolute terms and compared to the NMSSM. For each value of $m_{\tilde{t}}$ and $\mu$, $\tan\beta$ is determined independently for the twin and NMSSM benchmark models to obtain $m_h = 125 \GeV$. For the twin model, the parameter choices of $\lambda = 1.4$, $f=3v$, $m_A = 1.5 \TeV$, and $m^2_S = (1 \TeV)^2$ were chosen as an approximate best-case scenario for tuning given the perturbativity and Higgs coupling  constraints, which will be discussed respectively in the Appendix and Sec.~\ref{sec:couplingFits}. For the NMSSM we chose also a roughly optimal parameter point of $\lambda = 0.6$, $m_A = 0.8 \TeV$, and $m^2_S = (1 \TeV)^2$.

As discussed above, the improvement in tuning compared to the NMSSM at low stop masses is small due to the large value of $\tan\beta$ necessary to obtain the correct Higgs mass. However, at large stop masses the effective SUSY twin quartic becomes large and the degree of tuning remains better than $1\%$ out to $m_{\tilde t} \approx 3.5 \TeV$ and $\mu \approx 1 \TeV$. At this point the degree of tuning for the twin model is better by a factor of $\sim3.5$ than the NMSSM. There is also an unintuitive mild increase in tuning at small values of $\mu$ in the  SUSY twin model due to the structure of the RG equations for the singlet and Higgs soft masses.


\begin{figure}[t]
  \centering
  \includegraphics[width=.48\columnwidth]{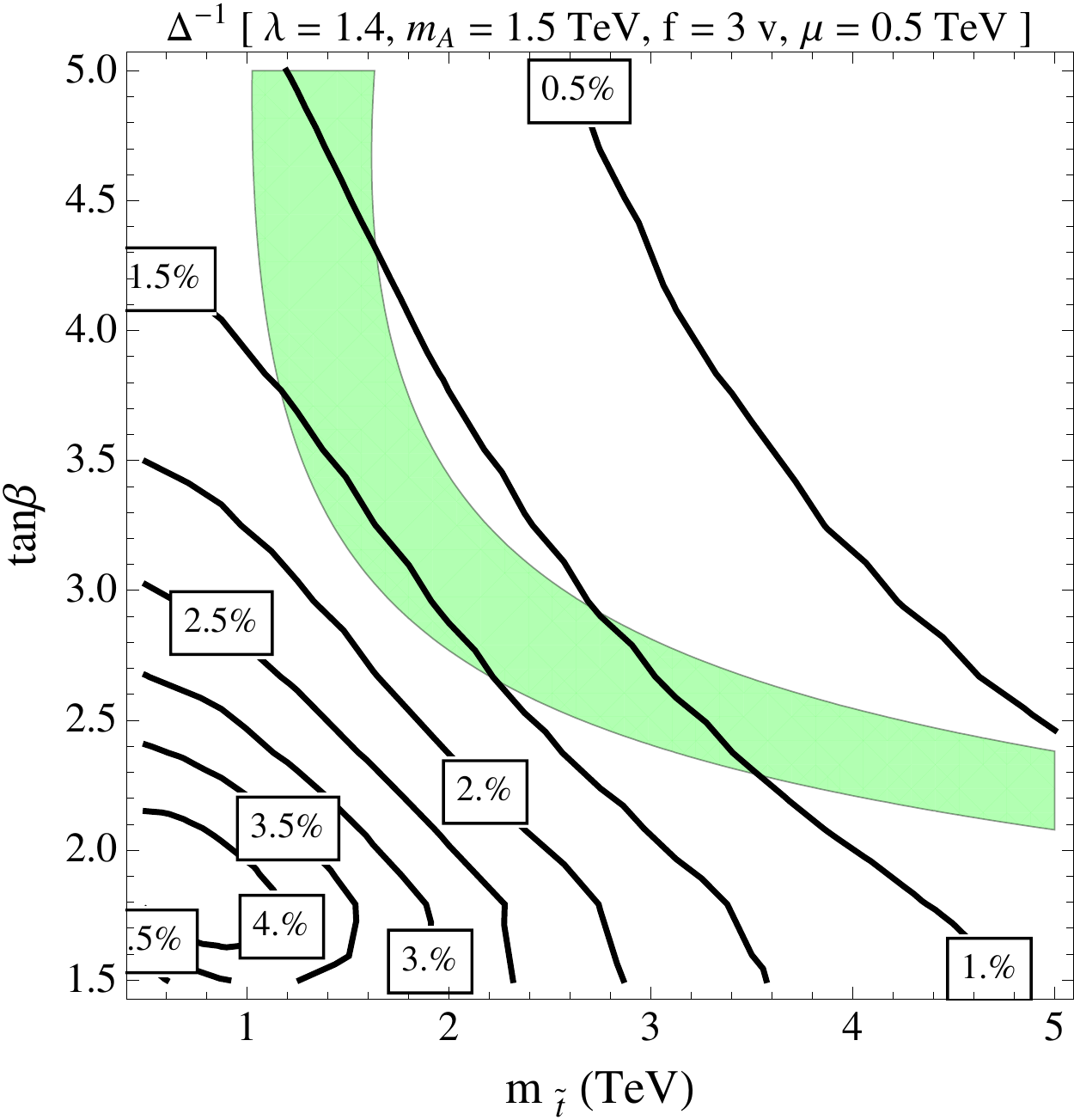}
  \caption{Tuning in the twin SUSY model with $\lambda = 1.4$, $f=3v$, $m_A = 1.5 \TeV$,  $m^2_S = (1 \TeV)^2$, $\mu = 0.5 \TeV$. The green shaded region is $123 \GeV < m_{h} < 127 \GeV$. }
    \label{fig:tuningHiggsMass}
\end{figure}

The consequences of the measured value $m_{h} \approx 125 \GeV$ on the tuning of the SUSY twin Higgs model are emphasized in Fig.~\ref{fig:tuningHiggsMass}. For this value of the Higgs mass, \emph{additional} U(4) breaking quartic couplings actually \emph{decrease} the tuning of the model by allowing the light Higgs mass to be obtained at smaller values of  $\tan \beta$. An important consequence is that the SUSY twin model is much more effective at reducing the tuning for stop masses of a few TeV, where the radiative contributions to the Higgs mass allow a small value of $\tan\beta$. This also raises the interesting possibility of decreasing the tuning at low stop masses by including extra tree-level U(4) breaking quartics. A simple example would be to expand the singlet sector to include independent singlets $S_A$ and $S_B$ coupling separately to the A and B sector Higgses to introduce NMSSM-like quartics. A modest value for the new singlet couplings $\lambda_{\cancel{\rm U(4)}} \sim 0.2-0.4$ could lift the Higgs mass to the measured value at low $\tan \beta$.  For example, for $m_{\tilde t}=1\TeV$, $\tan\beta=1.7$, and $\lambda=1.4$, we find that a tuning of better than $10\%$ can be obtained (a factor of $\sim3$ improvement over the NMSSM) and the Higgs mass can be accomodated with  $\lambda_{\cancel{\rm U(4)}} \sim 0.4$. For simplicity we do not include this non-minimal contribution to the Higgs mass in any of the following results unless otherwise noted.

\begin{figure}[t]
  \centering
  \includegraphics[width=.48\columnwidth]{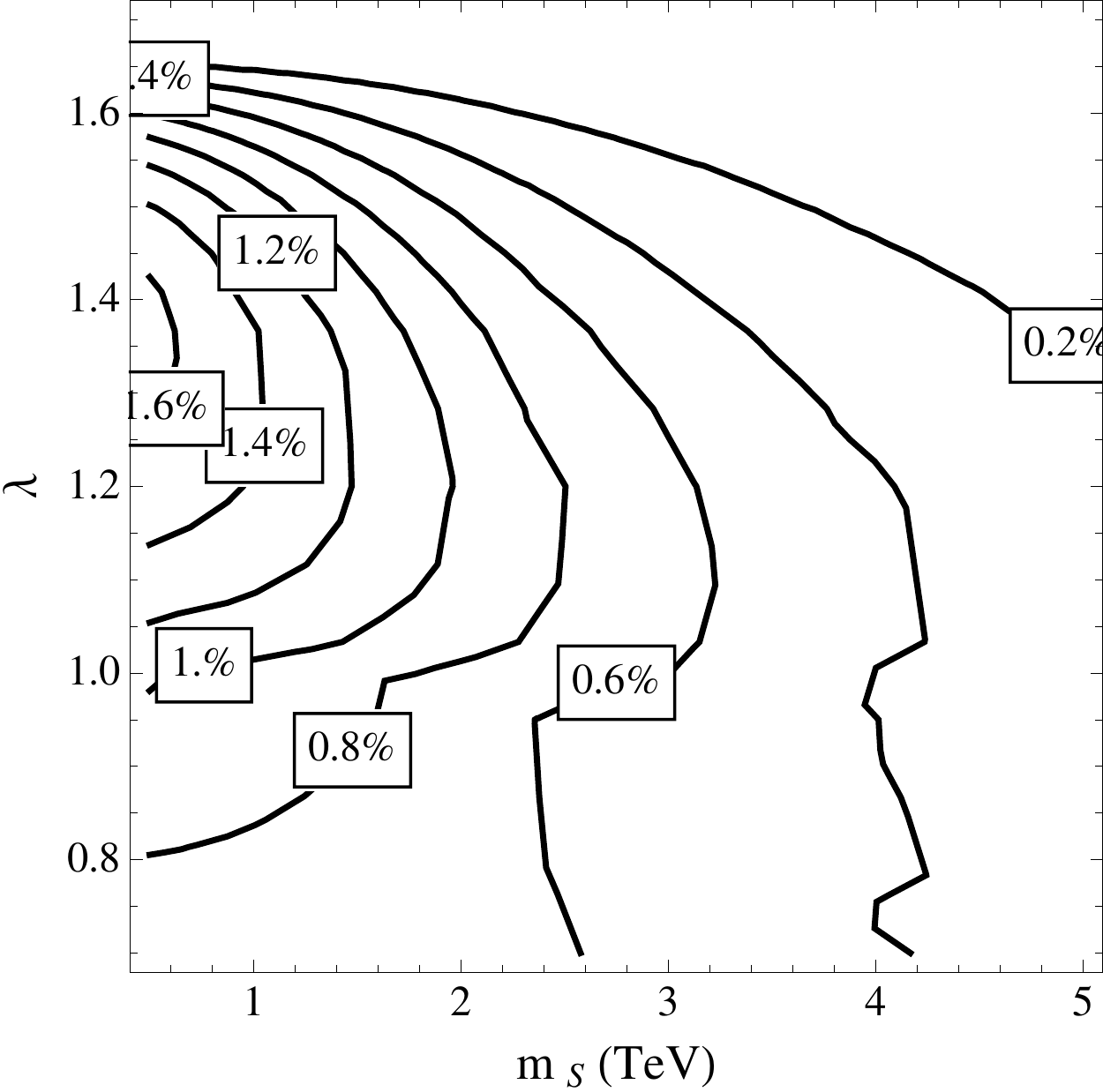}
  \caption{Tuning in the twin SUSY model as a function of $\lambda$ and $m^2_S$ with $f=3v$, $m_A=1.5 \TeV$, $m_{\tilde{t}}=2.0 \TeV$, and $\mu = 0.5 \TeV$.  At each point, $\tan\beta$ is determined to obtain $m_h = 125 \GeV$.}
    \label{fig:tuningSinglet}
\end{figure}

The soft mass of the singlet plays two important roles in determining the tuning of the twin SUSY model. First, it makes a contribution to the running of the Higgs masses which is important especially for large values of $\lambda$. The sensitivity  of the tuning to this effect is depicted in Fig.~\ref{fig:tuningSinglet}. For $\lambda \gtrsim 1.5$, the Landau pole becomes too close to the messenger scale and the contributions to the running from the singlet become large (see Appendix \ref{sec:lambdaUV} for further discussion of the Landau poles and UV completion of the singlet). For smaller values  $\lambda\sim1.2-1.5$, a $1 \TeV$ singlet starts to make contributions to the tuning comparable to a $2 \TeV$ stop. As small as possible value for the singlet soft mass is therefore desirable. 

On the other hand, as discussed in Sec.~\ref{sec:setup} the singlet soft mass must be considerably larger than $\mu$ and the supersymmetric singlet mass $M_S$  to obtain a large tree level quartic from the singlet sector. We have therefore chosen a benchmark value of $m^2_S = (1 \TeV)^2$, allowing moderately sized $\mu$ and $M_S$ terms while still generating a large quartic and not generating too large of a contribution to the Higgs soft masses. An interesting possiblity to circumvent this tension between radiative tuning and generating a tree-level quartic is to modify the singlet-Higgs sector to take the form of the Dirac NMSSM of Ref. \cite{Lu:2013cta}, but we do not study this possibility in detail.

While we have allowed values of $\lambda$ and $\tan\beta$ such that the singlet requires a UV completion above the SUSY breaking scale, we have assumed large enough values for $\tan\beta$ that the top Yukawa remains perturbative up to the GUT scale (see Appendix \ref{sec:lambdaUV}). It is interesting to sacrifice MSSM-like gauge coupling unification and consider how natural the SUSY twin Higgs model can be made if low scale Landau poles in both the singlet and top Yukawa couplings are permitted, with the assumption that a suitable fat-Higgs-like \cite{Harnik:2003rs} composite Higgs sector can provide a UV completion. The point $m_{\tilde t} \approx 3.5 \TeV$ and $\mu \approx 1 \TeV$ provides a useful benchmark. For $\lambda=1.4$ and $\tan\beta=2.6$, the correct Higgs mass is obtained with a tuning of $1\%$ and a Landau pole for the singlet near $\sim 500 \TeV$ that can be UV-completed consistent with gauge coupling unification. The tuning can be improved tenfold to $10\%$ for $\lambda=2$ and $\tan\beta=1.1$, which is a factor of 30 less tuned than the NMSSM benchmark for the same point. The cost of this decrease in tuning is that the singlet Landau pole is brought down to $\sim 50 \TeV$, and likewise the top Yukawa must be completed before the GUT scale. A version of the fat Higgs \cite{Harnik:2003rs}  could provide the necessary UV completion but appears incompatible with precision gauge coupling unification. The extra U(4) breaking quartics $\lambda_{\cancel{\rm U(4)}}\sim0.2$ must also be allowed to obtain the correct Higgs mass. Although this is an interesting possibility for dramatically reducing the fine-tuning in a (semi-)perturbative SUSY model, our primary interest in what follows will remain on the case where $\tan\beta$ is large enough that only the singlet requires UV completion and MSSM-like grand unification can occur.

\begin{figure}[t]
  \centering
  \includegraphics[width=.48\columnwidth]{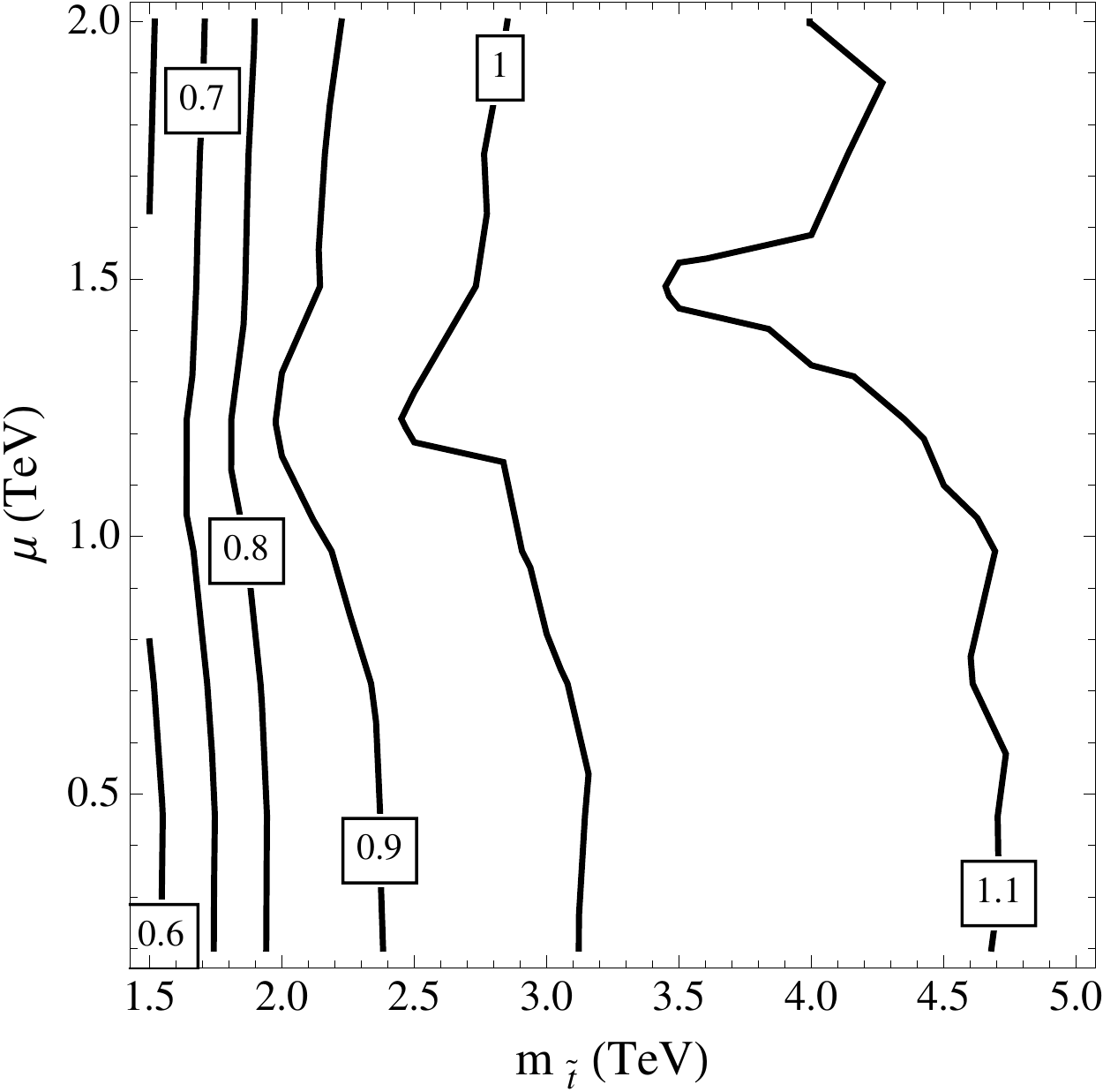}
  \caption{Ratio of tuning for twin SUSY models with ($f=3v$, $m_A=1.5 \TeV$) versus ($f=5v$, $m_A=2.0\TeV$). For both models $\lambda=1.4$ and $m^2_S = (1 \TeV)^2$.  At each point, $\tan\beta$ is determined independently for each of the models to obtain $m_h = 125 \GeV$. }
    \label{fig:tuningfCompare}
\end{figure}

In section \ref{sec:couplingFits} we will discuss the limits on $\frac{v^2}{f^2}$ from the observed couplings of the light Higgs state at the LHC. From Eq.~\ref{eq:TwinTuningApprox} we expect the total tuning to become roughly independent of $v/f$ in the limit of large $\mu$ or $m_{\tilde{t}}$ and $f^2 \gg v^2$. In fact,  because smaller $v^2/f^2$ allows the Higgs mass to be obtained at smaller values of $\tan\beta$, the tuning can be slightly reduced in this limit. Fig.~\ref{fig:tuningfCompare} demonstrates this behavior comparing the tuning at $f=3v$ to $f=5v$. For $m_{t} \lesssim 3 \TeV$ the $f=5v$ model is more tuned because the stop masses are not yet saturating the $f$-tuning, but for larger stop masses the $f=5v$ model accommodates the Higgs mass at smaller $\tan\beta$ and is slightly less tuned.

\subsection{An emergent \ZTwo}
\label{sec:emergent}

The crucial aspect of the twin Higgs mechanism is that the \ZTwo symmetry is realized in the gauge couplings, the large Yukawa couplings in the Higgs-top-singlet sector, and the soft SUSY breaking terms. On the other hand, sources of \ZTwo breaking in the Higgs potential are important to obtain a hierarchy in the A- and B-sector vevs, and \ZTwo breaking in the small Yukawa couplings is necessary to address cosmological complications as will be discussed in Sec.~\ref{sec:cosmology}. An interesting possibility is that the necessary \ZTwo symmetries of the large couplings are emergent in the IR while the smaller couplings reflect an $\mathcal{O}(1)$ breaking of the $\ZTwo$ in the UV superpotential. 

To be concrete, consider a UV model where the \ZTwo symmetry of the field content and gauge couplings is exact at $\Lambda_{\rm GUT}$, but the \ZTwo is broken by $\mathcal{O}(1)$ differences in couplings in the Higgs-Yukawa and Higgs-singlet sector.  The singlet must be UV completed to a composite state to allow large values of the IR coupling $\lambda$, as will be discussed in detail in Appendix~\ref{sec:lambdaUV}. The important detail for this discussion is that the IR couplings of both the A and B sectors Higgs to the composite singlet sector are governed by the same interacting fixed point with scaling dimensions set by the common strong sector. Therefore the low energy \ZTwo in the Higgs singlet couplings will emerge as long as sufficient time is spent in the interacting fixed point regime.  In models where the Higgses themselves are composite, the \ZTwo of the top Yukawa couplings can also emerge from the interacting fixed point. Even with an elementary Higgs-top sector, the low values of $\tan\beta$ preferred in the SUSY twin model  put the top Yukawa near the IR attractor value, making it insensitive to the value at $\Lambda_{\rm GUT}$ \cite{Schrempp:1996fb}. For example, for $\tan\beta\sim2.0$, $y_t(100\TeV)$ varies by only $5\%$ for $y_t(\Lambda_{\rm GUT})=0.5-2.0$ (disregarding the contributions to the running from the singlet sector, which will move the fixed point to a larger value of $\tan\beta$). This corresponds to $\lesssim10\%$ \ZTwo breaking in the stop contribution to the Higgs soft masses, which is consistent with the \ZTwo breaking necessary to create a hierarchy in vevs. The two-loop contributions of the top Yukawa to the gauge couplings leads to a negligible \ZTwo breaking in the gauge sector, and in a pure gauge mediation model the \ZTwo symmetry of the gauge couplings automatically leads to \ZTwo preserving SUSY breaking masses. If direct messenger-Higgs couplings are necessary, the \ZTwo symmetry in these coupling can emerge in the IR from similar attractor behavior.

The SUSY twin Higgs model therefore has the appealing property that the entire U(4) symmetry protecting the light Higgs state results from an IR \ZTwo symmetry of the Higgs-top-singlet-gauge sector which can itself emerge from a UV theory with $\mathcal{O}(1)$ breaking of the \ZTwo in the superpotential.

\section{Phenomenology}
\label{sec:pheno}

The low-energy phenomenology of the SUSY twin Higgs differs radically from conventional supersymmetric scenarios. Although the details of the sparticle spectrum require a complete model for supersymmetry breaking and mediation, it is clear that the stops and higgsinos can be significantly decoupled in the SUSY twin Higgs without increasing the tuning of the weak scale. Percent-level naturalness is consistent with stops at 3.5 TeV and higgsinos at 1 TeV, well beyond the reach of the 13/14 TeV LHC \cite{Gershtein:2013iqa}. In general, radiative corrections tie the mass of the gluino to within a factor of 2 of the stop mass, and gluinos in the range of 3.5-7 TeV likewise lie well beyond the reach of the 13/14 TeV LHC, although such heavy spectra would likely be accessible at an LHC energy upgrade \cite{Cohen:2013xda}. The avatars of double protection are likewise inaccessible at the LHC, since the fermionic top partner -- in the guise of the B-sector top quark -- is neutral under the Standard Model gauge groups and only pair-produced with minuscule cross section through the Higgs portal.

In the absence of conventional supersymmetric signals, the primary experimental indications of the SUSY twin Higgs come from the Higgs sector -- both in modifications of the couplings of the Standard Model-like Higgs, and in the multitude of additional states in the extended electroweak symmetry breaking sector.

\subsection{Higgs couplings}

The principal constraints on the SUSY twin Higgs arise from tree-level modifications to the couplings of the Standard Model-like Higgs, which we identify with the lighter CP-even neutral Higgs of the A-sector. The couplings of the SM-like Higgs are modified by both the usual SUSY mixing within the two A-sector Higgs doublets, as well as the mixing with B-sector Higgs doublets.\footnote{One could also look for NLO effects coming from loops of B-sector top quarks as in \cite{Craig:2013xia}, but these are typically subdominant to the tree-level coupling deviations.}

To the extent that we would like to constrain the SUSY twin Higgs parameter space with coupling measurements of the SM-like Higgs, the most interesting properties are those of the lightest CP even Higgs. In general, the matrix of mixings in the CP even Higgs sector is unenlightening, but we may capture the important parametrics by carrying out a perturbative expansion in the U(4) limit, $g^2 + g'^2\ll \lambda^2 $. Since we are interested in a relatively high scale of sparticles, we will also focus on the ``SUSY twin decoupling limit'' $\lambda^2 f^2 \ll m_A^2$, akin to the usual SUSY decoupling limit, $m_Z^2 \ll m_A^2$. Here $m_A^2 \equiv \frac{2 b}{\sin(2 \beta)}$ is a mass parameter that corresponds to the usual MSSM definition, as well as the mass of one physical pseudoscalar; it provides a convenient means of packaging results, and preserves the customary intuition that certain additional Higgs states decouple in the limit $m_Z^2 \ll m_A^2$.

 To leading nontrivial order in the expansion $g/ \lambda, m_Z / m_A, \lambda f / m_A \ll 1$, the four CP-even masses are $m_Z^2 \cos(2 \beta)^2 \left( 2 - \frac{2v^2}{f^2} \right), \, m_A^2 - \lambda^2 f^2, \, \lambda^2 f^2 \sin(2 \beta)^2,$ and $m_A^2 - \lambda^2 f^2 \sin(2 \beta)^2$. The first corresponds to the Goldstone mode, primarily identified with the A-sector light CP-even Higgs, while the remaining states are primarily identified with the A-sector heavy CP-even Higgs, B-sector light CP-even Higgs, and the B-sector heavy CP-even Higgs, up to inter-sector mixings of order $v/f$. With this in mind, in what follows we label the corresponding mass eigenstates $h_1 (\equiv h), H_1, h_2,$ and $H_2$, respectively, with $h$ identified with the recently-discovered SM-like Higgs. We adopt a similar nomenclature for the pseudoscalars $A_1, A_2$ and the charged Higgs pairs $H_1^\pm, H_2^\pm$.

The composition of $h$ in terms of the gauge eigenstates is very nearly what one would expect from the direct product of a supersymmetric 2HDM and the twin Higgs mechanism, viz.
\begin{eqnarray} 
h &\approx& \left[ \left( 1 - \frac{v^2}{2 f^2} \right) \sin \beta +  \frac{m_Z^2}{4 \lambda^2 f^2} \cos^2(2 \beta) \csc \beta \sec^2 \beta + 2 \frac{m_Z^2}{m_A^2} \cos^2 \beta \sin \beta  \cos(2 \beta)\right] {h^0_u}^A \\ \nonumber
&+& \left[ \left( 1 - \frac{v^2}{2 f^2} \right) \cos \beta   +  \frac{m_Z^2}{4 \lambda^2 f^2} \cos^2(2 \beta) \csc^2 \beta \sec \beta - 2 \frac{m_Z^2}{m_A^2} \cos \beta  \sin^2 \beta \cos(2 \beta)     \right] {h^0_d}^A \\ \nonumber
&+& \left[ \frac{v}{f} \sin \beta + \frac{m_Z^2}{4 \lambda^2 v f} \cos^2(2 \beta) \csc \beta \sec^2 \beta \right] {h^0_u}^B  + \left[ \frac{v}{f} \sin \beta + \frac{m_Z^2}{4 \lambda^2 v f}\cos^2(2 \beta) \csc^2 \beta \sec \beta \right] {h^0_d}^B 
\end{eqnarray}

Indeed, in the limit $g/ \lambda \to 0$ the mixing contributions from the SUSY 2HDM and the twin Higgs factorize such that the couplings of the Standard Model-like Higgs to vectors, top quarks, bottom quarks, and leptons are modified by an amount
\begin{eqnarray}\nonumber 
c_V &\approx& 1 - \frac{v^2}{2 f^2} - \frac{m_Z^4}{8 m_A^4} \sin^2(4 \beta) + \dots \\  \label{eq:higgscoup} 
 c_t &\approx& 1 - \frac{v^2}{2 f^2} + \frac{2 m_Z^2}{m_A^2} \cos^2\beta \cos(2 \beta) + \dots \\ \nonumber
   c_b = c_\tau &\approx& 1 - \frac{v^2}{2 f^2} - \frac{2 m_Z^2}{m_A^2} \sin^2 \beta \cos(2 \beta) + \dots
\end{eqnarray}
where $c_i \equiv g_{hii} / g_{h_{SM} ii}$. 

In addition to modifications of the Higgs couplings to Standard Model states, there is also generically an invisible width coming from decays of the Higgs to B-sector fermions, predominantly  $h \to b_B \bar b_B$, due to the $\mathcal{O}(v^2/f^2)$ mixing of $h$ with the B-sector CP-even Higgses. The B-sector bottom quark mass and couplings are fixed by the $Z_2$ symmetry, and so the partial width for $h \to b_B \bar b_B$ at leading order is 
\begin{equation} \label{eq:invis}
\Gamma(h \to b_B \bar b_B)  \approx \Gamma(h \to invis.) \approx \Gamma(h \to b \bar b) \tan^2(v/f) \left( \frac{1 - 4 \frac{m_b^2}{m_h^2} \frac{f^2}{ v^2} }{1 - 4\frac{ m_b^2}{m_h^2}} \right)^{3/2}
\end{equation}

The modified Higgs couplings and invisible width have two novel implications. The first is that the intrinsic $\mathcal{O}(v^2/f^2)$ tuning of the theory is set by measurements of Higgs couplings, much as in composite Higgs models. Since the precision of current Higgs coupling measurements in combination approaches $\mathcal{O}(10\%)$, this suggests $f \gtrsim 3 v$; we will make this statement more precise in the next subsection.

The second novel implication is that the invisible Higgs width is also of order $v^2 / f^2$, so that the invisible width of the Higgs directly probes the tree-level naturalness of the theory. Whereas the mass scale of higgsinos and top partners in the SUSY twin Higgs provides little concrete information regarding the naturalness of the weak scale, the invisible width provides an unambiguous indication.

\subsection{Coupling Fits}
\label{sec:couplingFits}

To establish the allowed range of both $v/f$ and the 2HDM mass scale $m_A$, we construct a combined fit to Higgs couplings using available data from both ATLAS and CMS searches at 7 and 8 TeV.\footnote{For fits to the related left-right twin Higgs model \cite{Chacko:2005un, Goh:2006wj} using various stages of LHC Higgs data, see \cite{Carmi:2012yp, Carmi:2012in, Liu:2013dma}.} To do so, we adopt the methods of \cite{Azatov:2012qz}. We construct a likelihood for each individual exclusive channel in \cite{Azatov:2012qz} using a two-sided gaussian whose mean is given by the experimental value of the signal strength modifier $\mu$ and whose width is given by the 1$\sigma$ errors on $\mu$. Where two-sided measurements are unavailable, we use an approximate gaussian likelihood constructed from the observed and expected limits.\footnote{Note that the channels in \cite{Azatov:2012qz} do not include direct limits on the Higgs invisible branching ratio from e.g. \cite{ATLAS:2013pma, CMS:2013bfa, CMS:2013yda}. To check the effects of the direct invisible branching ratio limit on the fit, we construct a single-channel likelihood using the numerical values of $-2 \log \mathcal{L}$ for the invisible branching ratio measurement in  \cite{ATLAS:2013pma}. The effects of invisible branching ratio limits \cite{CMS:2013bfa, CMS:2013yda} are very similar. The inclusion of this likelihood leads to an insignificant change in the best-fit region since the direct limit of ${\rm Br}(h \to {\rm invis.}) < 0.65$ is much weaker than the implicit limit in the  SUSY twin Higgs framework coming from measured branching ratios. We do not include this invisible branching ratio likelihood in our final fit due to uncertainties in the shape of the likelihood in \cite{ATLAS:2013pma} for low values of the invisible branching ratio. }

To determine the dependence of the signal strength on the relevant SUSY twin Higgs parameters $f, m_A,$ and $\tan \beta$, we use the techniques of \cite{Craig:2013hca} with the tree-level coupling modifiers in Eq.~\ref{eq:higgscoup} plus the invisible width in Eq.~\ref{eq:invis}. For simplicity, we take the limit $(g^2 + g'^2) / \lambda^2 \to 0$, for which the effects due to the SUSY 2HDM and twin Higgs sector approximately factorize. We have checked that this approximation to the Higgs couplings always agrees with full numerical results to within a few percent in the parameter regions of interest. We do not include any contributions from loops of superpartners, since the mass scale of superpartners is sufficiently high that these contributions are negligible.

\begin{figure}[t]
  \centering
  \includegraphics[width=.6\columnwidth]{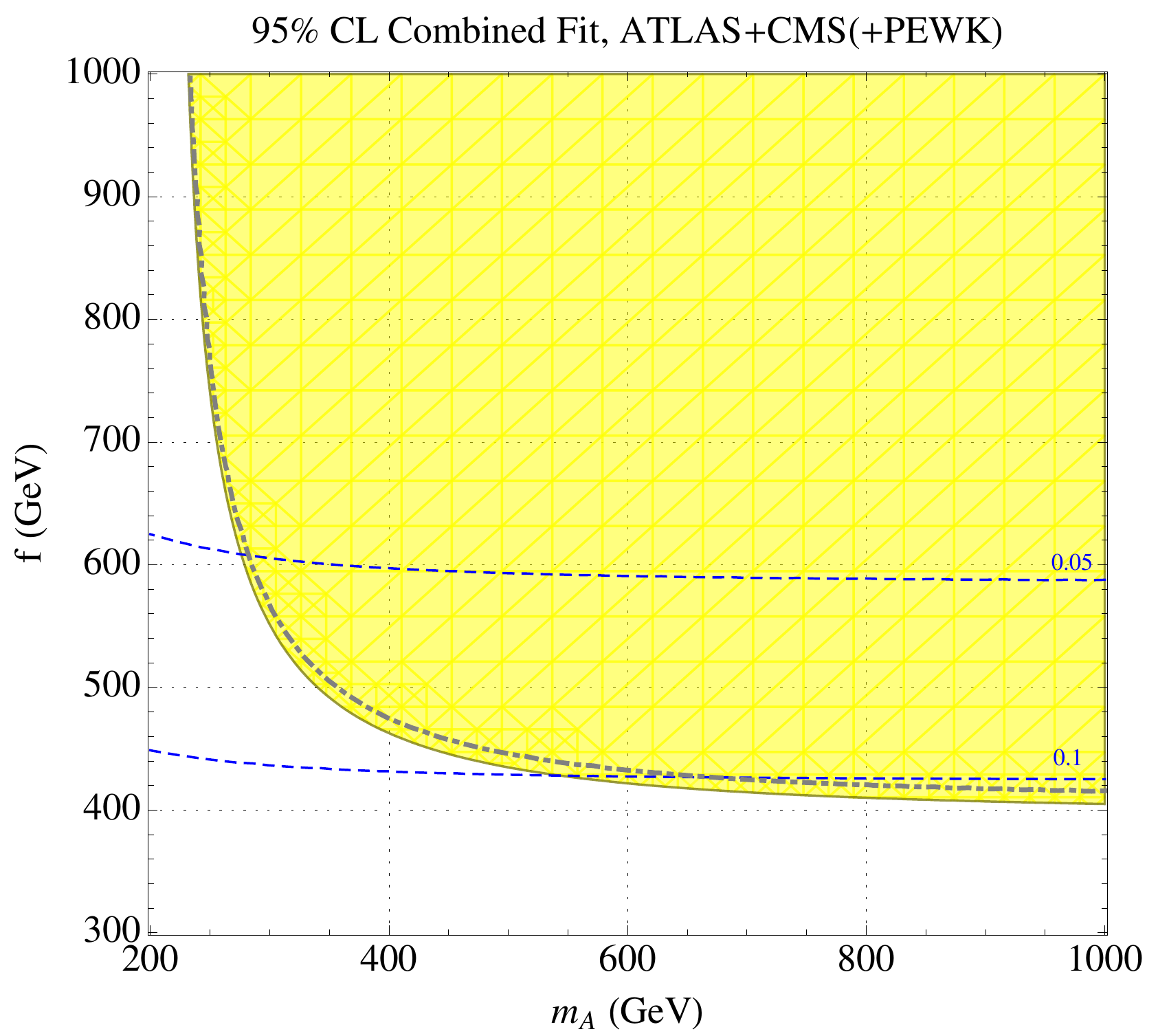}
  \caption{Coupling fit in the SUSY twin Higgs model as a function of $m_A$ and $f$ for the representative value of $\tan \beta = 2.5$ in the limit $(g^2 + g'^2) / \lambda^2 \to 0$. The fit procedure is described in the text. The yellow shaded region denotes the 95\% CL allowed parameter space defined by $-2 \Delta \ln \mathcal{L} < 5.99,$ not including precision electroweak constraints. The gray dot-dashed line denotes the edge of the 95\% CL allowed region including IR contributions to the $S$- and $T$-parameters, marginalizing over the $U$-parameter. The blue dashed lines indicate the contours of ${\rm Br}(h \to {\rm invis.}) = 0.05, 0.10$, respectively.}
    \label{fig:fit}
\end{figure}

Given these single-channel likelihoods, we construct a combined likelihood from the product of the single-channel likelihoods. To perform the fit, we fix the representative value $\tan \beta = 2.5$ and compute $-2 \Delta \ln \mathcal{L}$ in the $f, m_A$ plane relative to the best-fit point of $f, m_A \to \infty$. We denote the 95\% CL region by $-2 \Delta \ln \mathcal{L} < 5.99$ in this 2D plane.

It is well known in the case of composite Higgs models that the strongest constraint comes from the combination of Higgs coupling measurements and precision electroweak data, including the IR contribution to the $S$- and $T$-parameters from the modification of the SM-like Higgs couplings to vectors \cite{Azatov:2012bz}. As we will discuss in the next section, the situation is substantially improved in the SUSY twin Higgs model. For the sake of illustration, we also show the 95\% best fit region including precision electroweak limits on the IR contribution to the $S$- and $T$-parameters (Eq.~\ref{eq:1pewk}) for $\lambda = 1.5$, marginalizing over the $U$-parameter. This includes the leading constraints from electroweak precision tests on modified couplings of the SM-like Higgs. We do not include UV contributions to the $S$- and $T$-parameters, which depend on the details of the heavy Higgs spectrum but are numerically of the same order as the IR contributions and decouple as $m_A \to \infty$.

The coupling fit is shown in Fig.~\ref{fig:fit}, which illustrates that $f \gtrsim 3v$ is comfortably allowed by current coupling measurements (recall we work in units where $v \approx 174$ GeV and $f$ is similarly normalized), with a modest invisible branching ratio of up to $\sim 10\%$. The invisible branching ratio in the SUSY twin Higgs consistent with current coupling measurements is smaller than the allowed values found in e.g. \cite{Belanger:2013kya} because the invisible width scales similarly to the modifications of tree-level couplings, leading to a tighter constraint.

\subsection{Extended Higgs sector}

The extended twin Higgs sector offers a plethora of additional states in the Higgs sector, including three additional CP-even neutral scalars, two pseudoscalars, and two pairs of charged Higgses. For the most part, these additional degrees of freedom are kinematically decoupled. As discussed above, the masses of the the ``heavy'' CP-even scalars $H_1, H_2$ are $m_{H_{1}}^2 \sim m_A^2 - \lambda^2 f^2, m_{H_2}^2 \sim m_A^2 - \lambda^2 f^2 \sin^2(2 \beta).$ Both charged Higgs pairs have masses of order $m_{H_{1,2}^\pm}^2 \sim m_A^2 - \lambda^2 f^2$ with subleading splittings of order $\mathcal{O}(m_W^2, m_{W_B}^2)$. The pseudoscalars have masses $m_{A_1}^2 \sim m_A^2 - \lambda^2 f^2, m_{A_2}^2 \sim m_A^2$. Consequently, all of these states are typically $\gtrsim$ TeV with correspondingly low production cross sections at the LHC. Moreover, the additional Higgs states coming predominantly from the A-sector enjoy the usual decoupling properties of a SUSY 2HDM, with correspondingly small couplings to Standard Model gauge bosons and the SM-like Higgs $h$. Given the limited reach for narrow Higgs scalars $\gtrsim$ TeV, it seems unlikely that these degrees of freedom can be meaningfully probed at the LHC.

 However, the second-lightest CP-even neutral scalar $h_2$ may remain relatively light, with $m_{h_2} \approx \lambda f \sin (2\beta)$. It possesses a coupling to top quarks of $\mathcal{O}(v/f)$, so that it is produced at the LHC via gluon fusion with a cross section $\sigma(gg \to h_2) \approx (v/f)^2 \sigma(gg \to h_{SM}),$ where $h_{SM}$ is a Standard Model Higgs of equivalent mass. Consequently, the gluon fusion cross section can remain relatively large, $\mathcal{O}(1\; {\rm pb})$ at $\sqrt{s} = 14$ TeV for $m_{h_2} \sim 500$ GeV.
 
The decay of $h_2$ is likewise promising. Although in general $h_2$ couples to degrees of freedom in the B-sector, it possesses a relatively large trilinear coupling with the SM-like Higgs $h$, $\lambda_{h_2 h h} \approx \frac{m_{h_2}^2}{2 \sqrt{2} f}$. Consequently, the partial width $\Gamma(h_2 \to h h)$ grows as $\sim  m_{h_2}^3/f^2$ with no small numerical suppression, and indeed parametrically competes with the partial width into B-sector gauge bosons. For the range of $\lambda$ of interest, the two-body decays of $h_2 \to Z_B Z_B, W_B W_B$ are kinematically accessible, so that $\Gamma(h_2 \to Z_B Z_B, W_B W_B)$ and $\Gamma(h_2 \to hh)$ differ only by kinematic factors and degree-of-freedom counting. Both decays dominate over decays to B-sector top quarks -- which are kinematically inaccessible for $\lambda \lesssim 2$ -- and decays to lighter B-sector fermions and gauge bosons. Note that $h_2$ is not exceptionally wide for $m_{h_2} \lesssim$ TeV, since $\Gamma_{\rm tot}/m_{h_2} \sim \lambda^2/16 \pi^2 \ll 1.$

Thus ${\rm Br}(h_2 \to hh) \sim \mathcal{O}(0.1-0.4)$ over a wide range of masses with no strong suppression from decoupling, in stark contrast to the heavy Higgs of the MSSM.\footnote{The violation of conventional 2HDM decoupling intuition here stems from the fact that there are two separate vacuum expectation values in the extended Higgs sector and genuine decoupling also requires $\lambda \to 0$. The trilinear coupling still exhibits the necessary property that $\lambda_{h_2 hh} v \to 0$ in the appropriate alignment limit $v/f \to 0$.} This raises the tantalizing prospect of a resonant di-Higgs signal at the LHC \cite{Barbieri:2005ri, Bowen:2007ia, Craig:2012pu, Dolan:2012ac   } with cross sections of order $\sigma \cdot {\rm Br}(pp \to h_2 \to hh) \sim$ 10 $-$ 500 fb at $\sqrt{s} = 14$ TeV for $m_{h_2} \sim 500-1000$ GeV. This should be compared to the Standard Model Higgs pair production rate, $\sim 34$ fb at  $\sqrt{s} = 14$ TeV, and will be easier to distinguish from background due to the boosted kinematics and resonant production mode. Unlike conventional 2HDMs (SUSY or otherwise), there is no competitive signal from $h_2 \to WW,ZZ$ (i.e., the massive A-sector gauge bosons), due to the suppressed coupling of $h_2$ to Standard Model gauge bosons. This strongly motivates searches for resonant di-Higgs production over a wide range of heavy Higgs masses.

Alternately (or perhaps in conjunction with the $h_2 \to hh$ signal), one may look for vector boson fusion production or $Z$-associated production of $h_2$ followed by invisible decay into B-sector states (primarily $W_B, Z_B$). The production cross section times invisible branching ratio for e.g. $\sigma \cdot {\rm Br}(pp \to q q h_2 \to jj + invis.)$  should be of order 10-100 fb at $\sqrt{s} = 14$ TeV for $m_{h_2} \sim 500-1000$ GeV, and could provide strong validation of a signal in $h_2 \to hh$ or serve as an independent detection mode in its own right. Similar sensitivity to an invisibly-decaying heavy Higgs scalar should be available in associated production with a $Z$ boson. At present, the ATLAS invisible Higgs search \cite{ATLAS:2013pma} and the CMS invisible Higgs search \cite{CMS:2013bfa} present limits for heavy Higgs masses up to $m_H = 300, 400$ GeV, respectively, but could be meaningfully extended to higher masses.  Clearly, if the remaining SUSY states lie above 1 TeV, these novel Higgs signatures may be the most promising direct signal of the SUSY twin Higgs at the LHC.

\subsection{LHC search strategy}

As we have seen, the most promising signals of the SUSY twin Higgs include $\mathcal{O}(v^2/f^2)$ deviations in the tree-level couplings of the SM-like Higgs; a modest $\mathcal{O}(v^2/f^2)$ invisible branching ratio; resonant pair production of the SM-like Higgs from a heavier CP-even Higgs with a large trilinear coupling and $\mathcal{O}(v^2/f^2)$-suppressed gluon fusion production; and vector boson fusion and/or $Z$ associated production of the heavier CP-even Higgs followed by decay to invisible final states. To the extent that measurements of Higgs couplings and invisible width will attain at best $\mathcal{O}(10\%)$ precision at the LHC, this motivates searching for resonant pair production of the SM-like Higgs and extending Higgs invisible width searches beyond $m_H = 300-400$ GeV. Discovery of either process would strongly motivate construction of a Higgs factory to further test for tree-level coupling deviations and a modest Higgs invisible width.

\section{Ancillary constraints}
\label{sec:limits}

\subsection{Precision electroweak and flavor}

In contrast to composite Higgs models, the precision electroweak corrections in the SUSY twin Higgs are all calculable and, by construction, quite small. A key advantage, even with respect to other natural models in which the Higgs is a goldstone boson, is the inertness of the heavy B-sector gauge bosons with respect to A-sector gauge bosons and fermions. There are therefore no tree-level contributions to the $S$ and $T$ parameters. This avoids the largest corrections to precision electroweak observables present in, e.g., little Higgs models.

Unsurprisingly, here the extended Higgs sector is the principal source of new contributions to precision electroweak observables. In general these contributions are completely consistent with current limits on $S$ and $T$. For the sake of simplicity and clarity, we'll restrict ourselves to a brief discussion of precision electroweak contributions in the limit $v \ll f \lesssim m_A$, in which case the Higgs sector consists of the light SM-like Higgs, a second CP-even Higgs $h_2$ around $m_{h_2} \sim \lambda f$, and the remaining Higgs scalars $H_{1,2}, A_{1,2}, H^\pm_{1,2}$ clustered around a common mass $m_A \gtrsim$ TeV. In this limit, we can simply treat the largest corrections to Standard Model expectations from the coupling deviations of $h$ and additional contributions of $h_2$. In the $m_h, m_{h_2}$ sector, the additional contributions to the $S$ and $T$ parameters -- {\it beyond} the usual contribution from a Standard Model Higgs of mass $m_h$ -- are given by
\begin{equation} \label{eq:1pewk}
\Delta S \approx \frac{1}{6 \pi} \left( \frac{v}{f} \right)^2 \log \left( \frac{m_{h_2}}{m_h} \right) \hspace{1cm} \Delta T \approx - \frac{3}{16 \pi \cos^2 \theta_W} \left( \frac{v}{f} \right)^2 \log \left( \frac{m_{h_2}}{m_h} \right)
\end{equation}

This captures the leading correction to the $S$ and $T$ parameters from variations in the couplings of the SM-like Higgs in the limit $m_A \to \infty$ and is quite small for the parameter range of interest. Additional corrections arise from the remaining Higgs scalars $H_{1,2}, A_{1,2}, H^\pm_{1,2}$ at the scale $m_A \gtrsim$ TeV. However, as in the MSSM, the these additional states decouple with increasing $m_A$; in particular the sectors $(H_1, A_1, H_1^\pm)$ and $(H_2, A_2, H_2^\pm)$ are approximately degenerate so that electroweak corrections are small. In the limit $g,g' \to 0$, the $(H_1, A_1, H_1^\pm)$ sector is exactly degenerate, and corrections from this sector to $S$ and $T$ vanish; for nonzero $g, g'$ this leads instead to the customary MSSM-like contributions that are strongly suppressed by $\mathcal{O}(m_Z^2 / m_A^2)$ in the regime of interest. Corrections from the $(H_2, A_2, H_2^\pm)$ sector are additionally suppressed by a factor of $v^2/f^2$ due to the smallness of mixing with the A sector, but in the $g,g' \to 0$ limit nonzero splitting between $H_2, A_2,$ and $H_2^\pm$ persists. Expanding the appropriate loop functions (e.g., \cite{Baak:2011ze}),  in the limit $m_A^2 \gg \lambda^2 f^2$, the leading contributions to $S$ and $T$ from the $(H_2, A_2, H_2^\pm)$ sector are parametrically of order
\begin{equation} \label{eq:2pewk}
\Delta S \approx \frac{1}{16 \pi} \frac{\lambda^2 v^2}{m_A^2} \hspace{1cm} \Delta T \approx \frac{1}{48 \pi} \frac{\lambda^2}{g^2 s_W^2} \frac{\lambda^2 f^2}{m_A^2}.
\end{equation} 
There are also contributions from loops involving one scalar from each sector, but these share an overall suppression factor of  $\mathcal{O}(v^2/f^2)$ due to mixing, as well as a similar magnitude of mass splitting between states, leading to corrections of the same order as Eq.~\ref{eq:2pewk}. Taken together, the corrections to $S$ are insignificant, while the corrections to $T$ are typically numerically of the same magnitude as those in Eq.~\ref{eq:1pewk}, and both show the expected decoupling as $m_A \to \infty$. However, the corrections to $T$ have the potential to generate (mild) tension with precision electroweak limits if $m_A \sim \lambda f \lesssim$ TeV, though this depends in detail on the extended Higgs sector and is readily susceptible to cancellations. Finally, corrections from the remainder of the sparticle spectrum are parametrically small unless there is substantial mixing in the squark sector.

There are no pernicious new sources of flavor violation in the SUSY twin Higgs beyond those usually encountered at one loop. In particular, the extended Higgs sector automatically satisfies the Glashow-Weinberg condition \cite{Glashow:1976nt} due to a combination of holomorphy and gauge invariance, guaranteeing the absence of new tree-level contributions to flavor-changing neutral currents. At one loop, the decoupling of charged Higgs states protects against prohibitive contributions to, e.g., $b \to s \gamma$. Although sfermions may all be in the multi-TeV range, this alone is insufficient to suppress one-loop FCNC in the presence of large flavor-violating soft masses, and so the usual solutions to the supersymmetric flavor problem are still required.

\subsection{Cosmology}
\label{sec:cosmology}

The cosmology of mirror twin Higgs models has been discussed in detail in refs. \cite{Chacko:2005pe,Barbieri:2005ri,Chang:2006ra}, and for mirror models in general in refs. \cite{Berezhiani:1995am, Berezhiani:2005ek, Foot:2004pa}. Here we review briefly the important constraints from light degrees of freedom and dark matter abundance.

The principal cosmological constraints on mirror twin Higgs models are the CMB and BBN bounds on extra light degrees of freedom, most stringently the recent Planck result $N_{\rm eff} = 3.30\pm0.27$ \cite{Ade:2013zuv}. The light Higgs state keeps the A and B sector in thermal equilibrium down to temperatures $T_{\rm eq}\sim \mathcal{O}(1\GeV)$ \cite{Chang:2006ra}. In the limit of an exact \ZTwo symmetry there is an unacceptably large contribution to $N_{\rm eff}$ if the reheating is symmetric or $T_{\rm RH} \gtrsim T_{\rm eq}$. 

The possibility of asymmetric reheating is discussed in refs. \cite{Berezhiani:1995am,Berezhiani:2005ek,Foot:2004pa}. An alternative solution in a high-T symmetric reheating scenario is to include \ZTwo breaking contributions to the B-sector Yukawas to lift the light quarks and charged leptons above $T_{\rm eq}$ \cite{Chacko:2005pe,Barbieri:2005ri,Chang:2006ra}. This can be a hard breaking in the flavor sector or spontaneous breaking from asymmetric vevs of flavon fields; the small Yukawa couplings feed into the RG of the Higgs-top-gauge sector at acceptably small levels and do not modify the tuning. If only the B-sector photons, gluons and neutrinos remain below $T_{\rm eq}$, then in the absence of entropy production in the QCD phase transition $\Delta N_{\rm eff}\sim1.4$ \cite{Barbieri:2005ri}. This may be reduced to comfortably within bounds if the A-sector QCD phase transition involves entropy production not present in the B-sector transition (due to the presence of light quarks in the A-sector), or if the B-sector QCD phase transition is raised above $T_{\rm eq}$ \cite{Chacko:2005pe,Barbieri:2005ri}. The tension may be further ameliorated if B-sector gauge groups are spontaneously broken (via, e.g., tachyonic B-sector soft masses giving rise to charge- and color- breaking minima).

In mirror twin Higgs models, the B-sector baryon and lepton number are independently conserved and can lead to stable relics. If a baryon asymmetry is generated in the B-sector, the lightest B-baryon, which may be charged or neutral, can be the DM and naturally link the DM and SM baryon abundance \cite{Barbieri:2005ri}. The lightest B-sector charged lepton can also make up a component of the DM if its decay to a charged B-pion and neutrino is kinematically forbidden.  The phenomenology of charged baryonic or leptonic B-sector DM components depends in detail on whether or not the B-sector U(1) is broken and on the spectrum of hadronic and nuclear states.  In the SUSY twin Higgs model, the lightest superpartner can also provide a dark matter candidate. The R-parity of the A and B sectors is shared, and a neutralino LSP can be a mixture of A and B-sector states. The larger B-sector Higgs vev can naturally lead to an LSP with primarily B-sector components, suppressing standard direct and indirect detection signals.

\section{Conclusions}

We have shown that a minimal supersymmetric completion of the mirror twin Higgs model yields MSSM-like gauge coupling unification, a naturally light SM-like Higgs, and small corrections to electroweak precision and flavor observables.  The level of tuning of this model is comparable to the NMSSM with a superpartner mass scale half as large, and the observed $\sim 125 \GeV$ mass of a SM-like Higgs state is consistent with a percent-level tuned spectrum of superpartners likely unobservable at both the 13/14 TeV LHC and a $\sim1\TeV$ linear collider. Provided additional U(4)-breaking quartics, a spectrum with superpartners at current LHC limits is consistent with tuning at the $10\%$ level. Furthermore, if we discard the requirement of perturbative MSSM-like gauge coupling unification, a Higgs compositeness scale of $\sim50\TeV$ allows $10\%$-level tuning with superpartners entirely out of reach of the LHC. 

With the superpartners in these models out of reach, the most promising collider signals come from the Higgs sector. The mixing of the lightest Higgs with the mirror sector is proportional to the hierarchy of vevs $\frac{v}{f}$, and constraints on the Higgs couplings translate into a direct and unambiguous constraint on the fine-tuning of the model, $\Delta^{-1} < \frac{2v^2}{f^2}$. Already the measurements of the couplings of the SM-like Higgs state at the LHC and precision electroweak measurements require a hierarchy in vevs $\frac{v}{f}  \lesssim \frac{1}{3}$, and few-percent-level measurements of Higgs couplings at the 13/14 TeV LHC will put more stringent limits on this model.  The Higgs coupling limits we derived in the supersymmetric decoupling limit also apply equally well to any low-energy effective twin Higgs theory. While most of the extra Higgs states can easily be decoupled, the next-to-lightest CP-even Higgs state  typically remains within reach of the 13/14 TeV LHC and has large branching ratios to both the striking di-Higgs channel and invisible final states.  Just as the discovery of the light SM-like Higgs determined the size of the effective quartic self-coupling and made concrete the natural scale for physics cutting off the top quark contribution to the Higgs mass, measuring the mass of the next lightest CP-even Higgs state in the mirror twin model will point to the natural scale for the superpartners in the twin SUSY model. The presence of this light state is also an important signal that the twin mechanism is perturbatively realized, rather than resulting from compositeness at the scale of a few TeV.

The SUSY twin Higgs, like any model involving a ``double-protection" solution to the hierarchy problem, clearly presents a challenge from the point of view of UV model-building and parsimony. Compared to composite models, this issue is somewhat alleviated for the SUSY twin Higgs model; here the full approximate U(4) symmetry emerges accidentally from a smaller \ZTwo symmetry which can originate far in the UV and may in fact be partially emergent at low energies.  As a minimal supersymmetric extension of the twin Higgs, the model we have presented is also considerably more appealing from this point of view  than earlier efforts at supersymmetrizing the twin Higgs \cite{Chang:2006ra,Falkowski:2006qq}.   
If null results in searches for superpartners persist at the 13/14 TeV LHC, understanding in detail the signatures of models like the SUSY twin Higgs -- which trade parsimony for decreased fine-tuning -- will become crucial to interpreting the role of naturalness as a predictive principle for the next generation of collider, dark matter, and low-energy precision experiments.

There are many possible avenues for further study. In this paper we have focused on the low-energy phenomenology without committing to a detailed model for supersymmetry breaking. It would be interesting to investigate mediation mechanisms giving rise to the appropriate combination of U(4)-symmetric and U(4)-breaking soft terms, perhaps via gauge mediation with suitable Higgs-messenger couplings. A SUSY-breaking mechanism that gives rise to tachyonic scalars in the B sector would be attractive from the perspective of cosmology, where spontaneous breaking of B sector gauge symmetries helps to alleviate constraints from $N_{eff}$. We have also not discussed dark matter candidates in detail, but the super-abundance of dark matter candidates in the SUSY twin Higgs model could give rise to a number of interesting scenarios that merit further study. Finally, while we have presented parametric estimates for the rates of resonant di-Higgs and invisible heavy Higgs production, a detailed study of these signals and their prospects for LHC discovery would be worthwhile.

\subsection*{Acknowledgments}
We thank Masha Baryakhtar, Zackaria Chacko, Savas Dimopoulos, Tony Gherghetta, Roni Harnik, and John March-Russell for collaboration at various early stages of this work; Andrew McLeod, Marco Serone, David Shih, and Matt Strassler for useful conversations; and Jamison Galloway for enlightening discussions about Higgs coupling fits. We particularly thank  Jamison Galloway and Roni Harnik for commenting on a draft of the manuscript. NC is supported by the DOE under grants DOE-SC0010008, DOE-ARRA-SC0003883, and DOE-DE-SC0007897, and acknowledges the hospitality of the Stanford Institute for Theoretical Physics and ERC grant BSMOXFORD no. 228169.  KH is supported by an NSF Graduate Research Fellowship under Grant number DGE-0645962 and by the US DoE under contract number DE-AC02-76SF00515, and was partially supported by ERC grant BSMOXFORD no. 228169.

\appendix

\section{UV completion and $\lambda$}
\label{sec:lambdaUV}

We have seen that the naturalness of the SUSY twin Higgs model is improved for larger values of $\lambda$, which raises the prospect of hitting a Landau pole in $\lambda$ beneath the unification scale. A complete model that preserves the suggestive IR indications of gauge coupling unification should therefore include a suitable UV completion for $\lambda$.

\begin{figure}[t]
  \centering
  \includegraphics[width=.48\columnwidth]{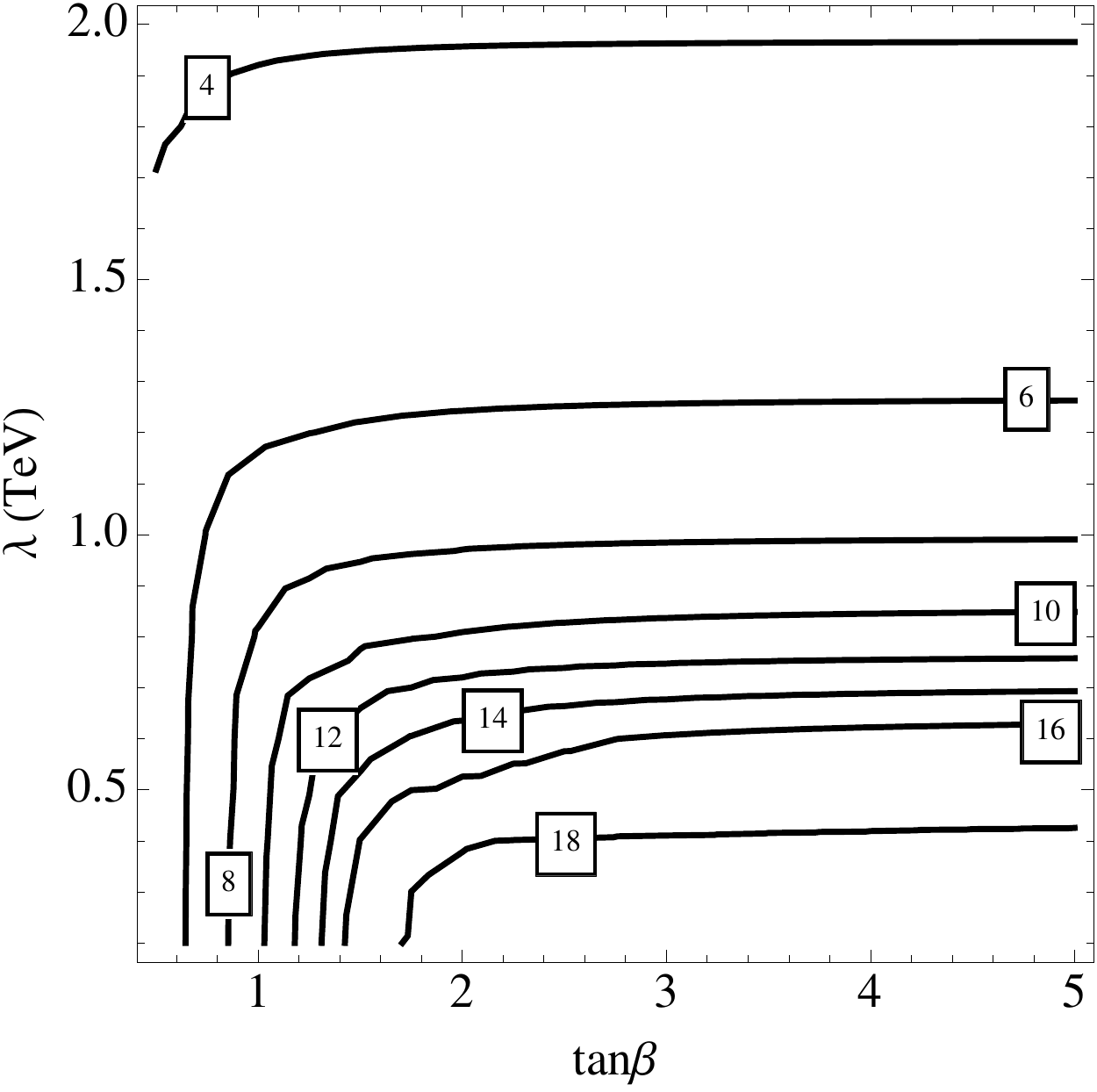}
  \caption{The approximate scale $\Lambda$ of the Landau pole in $\lambda$, defined by $\lambda(\Lambda) = \sqrt{4 \pi}$, as a function of $\lambda(1 \; {\rm TeV})$ and $\tan \beta$. Contours denote $\log_{10}(\Lambda / {\rm GeV}).$}
    \label{fig:landaupole}
\end{figure}

Although we are already accustomed to UV completions in the NMSSM for $\lambda({\rm TeV}) \gtrsim 0.7$, the twin Higgs $\lambda$ coupling hits a Landau pole faster than its NMSSM counterpart due to a larger $\beta$ function,
\begin{eqnarray}
\beta_\lambda ({\rm Twin}) = \frac{6 \lambda^3}{16 \pi^2}  + \dots \\
\beta_\lambda ({\rm NMSSM}) = \frac{4 \lambda^3}{16 \pi^2}  + \dots 
\end{eqnarray}
which causes $\lambda$ to increase faster in the UV. At one loop, the scale of the Landau pole depends on the weak-scale value of $\lambda$ and on $\tan \beta$ (via dependence on the top Yukawa). In Fig.~\ref{fig:landaupole} we show the approximate location of the Landau pole in $\lambda$ for the SUSY twin Higgs as a function of $\lambda({\rm TeV})$ and $\tan \beta$. From Fig.~\ref{fig:landaupole} it's clear that there is generically a Landau pole well below the GUT scale for the values of $\lambda({\rm TeV}) \gtrsim 1$ favored by naturalness. 

Although there are various possible approaches to UV completing the Landau pole in $\lambda$, the Slim Fat Higgs \cite{Chang:2004db} provides an attractive candidate insofar as it does not require a large amount of additional matter charged under the A- and B-sector gauge groups. The essential idea of the Slim Fat Higgs is that the singlet $S$ emerges as a meson of an $SU(N_c)$ gauge group that is deflected from an interacting fixed point to an s-confining fixed point by a mass term for some number of flavors. Concretely, in the UV we introduce an $SU(N_c)$ gauge group with $SU(N_c)$ (anti)fundamentals $\phi (\phi^c)$ that are SM singlets;  $SU(N_c)$ (anti)fundamentals $X (X^c)$ that are both A- and B-sector electroweak doublets with a Dirac mass $M_D$; $SU(N_c)$ (anti)fundamentals $\tilde X (\tilde X^c)$ with a Dirac mass $M_T$ that partner with $X, X^c$ to fill out complete A- and B-sector $SU(5)$ unified multiplets; and a number of additional $SU(N_c)$ (anti)fundamentals neutral under both A- and B-sector groups. In the twin version of the Slim Fat Higgs, the $X, \tilde X$ collectively account for $\delta N_f = 10$ flavors (i.e., a fundamental + antifundamental pair of both A-sector and B-sector $SU(5)$). The UV theory also includes superpotential couplings of the form $W \supset \lambda_1 \phi H_u X^c + \lambda_2 \phi^c H_d X$. At the scale $M_D$, the fields $X, X^c$ are integrated out to generate the effective operator
\begin{equation}
W \to - \frac{\lambda_1 \lambda_2}{M_D} \phi \phi^c H_u H_d
\end{equation}
At a scale $\Lambda$ at or beneath the scale $M_D$, the theory flows to an s-confining fixed point where $S \sim (\phi \phi_c)$, and we identify $\lambda(\Lambda) \equiv \lambda_1 \lambda_2 \frac{\Lambda}{M_D}$. The challenge is to construct a theory that is at an interacting fixed point at high energies (to retain UV completeness, and to guarantee that $\lambda_{1,2}$ are sufficiently large) and flows to an s-confining fixed point when $X, X^c, \tilde X, \tilde X^c$ are integrated out, while keeping $N_c$ small enough to avoid introducing too much matter charged under the Standard Model gauge groups. In the conventional Slim Fat Higgs model, the unique solution to these constraints was $N_c = 4$. In the twin Slim Fat Higgs model, the doubling of Standard Model gauge groups means that the $X, \tilde X$ account for twice the number of flavors under $SU(N_c)$, and the requirements of asymptotic freedom in the UV, s-confinement in the IR, and the avoidance of SM Landau poles cannot be simultaneously satisfied. However, there is a small modification of the Slim Fat Higgs model that suffices.

\begin{figure}[t]
  \centering
  \includegraphics[width=.38\columnwidth]{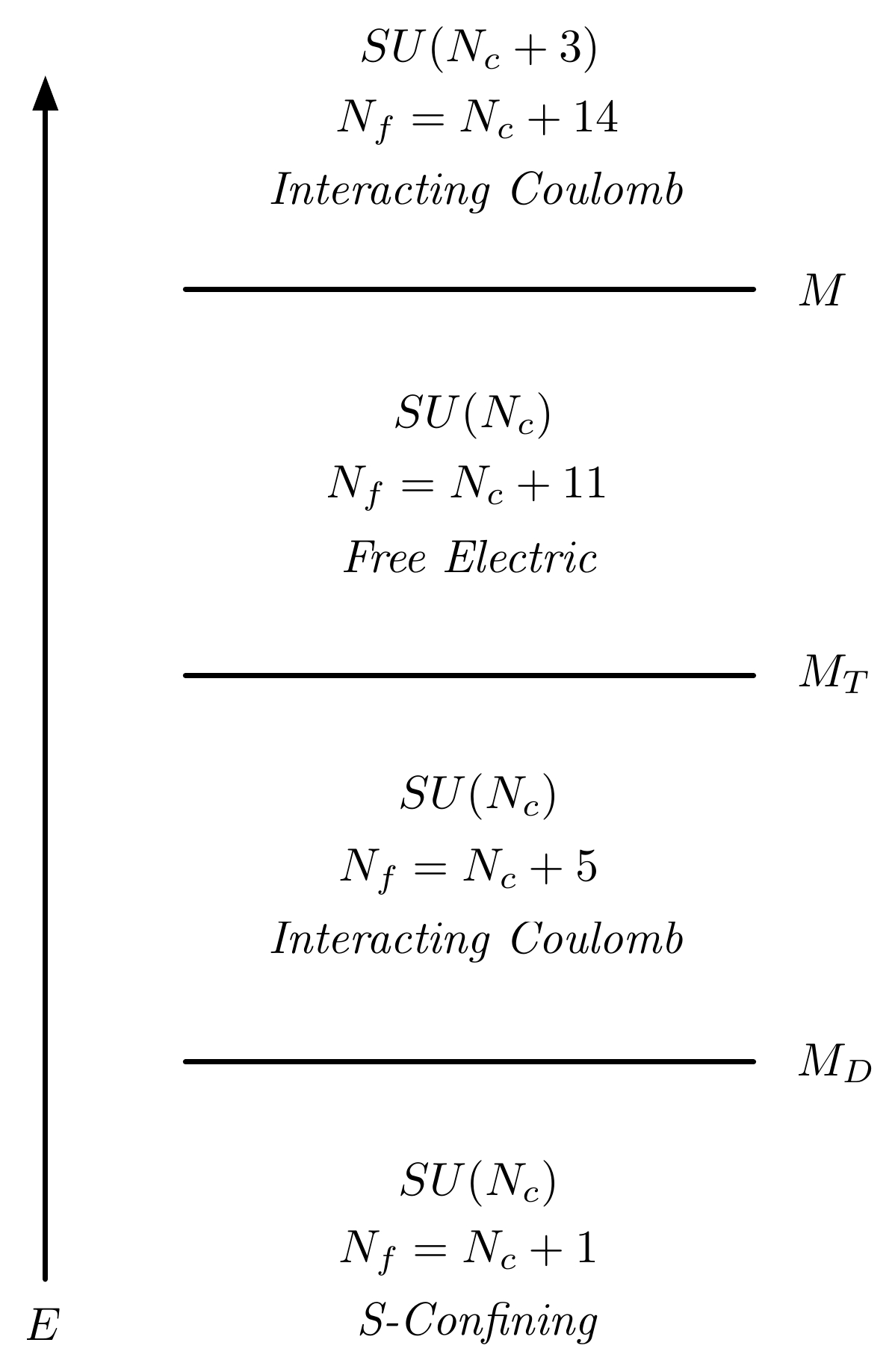}
  \caption{Cartoon of the scales and mass flow for the Slim Fat twin Higgs UV completion of $\lambda$.  }
    \label{fig:uvcompletion}
\end{figure}

A cartoon of the UV completion is shown in Fig.~\ref{fig:uvcompletion}. We begin with an $SU(N_c + 3)$ theory at high energy with $N_f = N_c+14$ flavors, which is asymptotically free for small ($N_c = 3,4,5$) values of $N_c$. This ensures that the theory remains under control in the UV. At the scale $M$, this theory is Higgsed to $SU(N_c)$ when three flavors acquire vevs and masses of order $M$. The remaining light flavors now transform as ${\bf N_c} + 3 \times {\bf 1}$, i.e., fundamental flavors plus singlets. These singlets can be given masses of order $\sim M$ by pairing with elementary singlets $\Sigma$ via superpotential interactions $W \supset \Sigma Q \tilde Q$. Thus the theory below $M$ now consists of an $SU(N_c)$ gauge theory with $N_f = N_c + 11$ fundamental flavors, which in general is in a free electric phase. This pattern of Higgsing ensures that the theory is asymptotically free in the UV, but also that there are only $N_c$ additional fundamental + antifundamentals charged under the Standard Model gauge group beneath the scale $M$. The rest of the story proceeds as with the usual Slim Fat Higgs; at the scale $M_T$, the triplets $\tilde X, \tilde X^c$ are integrated out, the theory has $N_f = N_c+5$ flavors and is generally back in an interacting Coulomb phase down to $M_D$, the scale where the doublets are integrated out and the theory s-confines. Since the theory is in an interacting phase above the scale $M$ and between $M_T$ and $M_D$, this generally ensures $\lambda_{1,2}$ are sufficiently large to offer a plausible UV completion for $\lambda$.

The primary constraint on $N_c$ arises from Landau poles in the A- and B-sector gauge couplings, since the $X, \tilde X$ fields collectively account for $N_c$ fundamental+antifundamental pairs under $SU(5)_A$ and $SU(5)_B$. In order to avoid A- and B-sector Landau poles, we would like $N_c$ to be as small as possible, but it should also be large enough that the theory is deep in the interacting Coulomb phase between $M_T$ and $M_D$, where the fixed point anomalous dimension  $\gamma_*$ of $\phi, \phi^c$ controls the size of $\lambda_{1,2}$. We may estimate the fixed point value of $\lambda_{1,2}$ in the weak coupling limit as in \cite{Chang:2004db}. The perturbative RGEs for $\lambda_{1,2}$ are 
\begin{equation}
\frac{d \lambda_{1,2}}{dt} = (N_c+3) \frac{\lambda_{1,2}^3}{16 \pi^2} + \gamma_* \lambda_{1,2} + \dots
\end{equation}
and so in the weak coupling approximation the fixed point value of the couplings above the scale $M_D$ is $\lambda_{1,2*} \approx 4 \pi \sqrt{|\gamma_*| / (N_c+3)}$.  Thus the NDA estimate for $\lambda$ at the confinement scale $\Lambda$ is 
\begin{equation}
\lambda(\Lambda) \approx \sqrt{N_c} \frac{\lambda_1 \lambda_2}{4 \pi} \frac{\Lambda}{M_D} \approx 4 \pi \frac{\sqrt{N_c}}{N_c + 3} |\gamma_*|
\end{equation} 
assuming $\Lambda \sim M_D$. For $N_c = 3,4,5$, we have $\gamma_* = -1/8, -1/3, -1/2$ and so $\lambda(\Lambda) \approx 0.45, 1.20, 1.76$. Needless to say, this is only an estimate due to the presence of additional incalculable $\mathcal{O}(1)$ factors, but it suggests that $N_c = 4,5$ both provide suitable UV completions for the values of $\lambda({\rm TeV})$ under consideration. 

The UV completion of the singlet sector can also have an important effect on the running of the top Yukawa coupling through potentially large contributions to the Higgs coupling, especially in the interacting Coulomb phase \cite{Hardy:2012ef}. The one-loop beta function for $y_t$ at the naive fixed point of the IR interacting Coulomb phase is
\begin{equation}
\frac{d y_t}{d t} \sim \begin{cases} 0.2 y_t & N_c = 4 \\ 0.3 y_t & N_c=5 \end{cases}.\label{eq:yt}
\end{equation}
With the low values of $\tan\beta$ preferred by the Higgs mass and tuning, Eq.~\ref{eq:yt} suggests that the theory can not remain near the interacting fixed point very long without increasing $y_t$ sufficiently to run non-perturbative before the GUT scale. Although this introduces some tension with the notion of natural MSSM-like gauge coupling unification in this model, we note that a value near the perturbativity bound is a natural expectation in models where the low energy $Z_2$ symmetry is partially emergent as discussed in Sec.~\ref{sec:emergent}.

\footnotesize
\setstretch{1.0}
\bibliography{TwinSUSY}
\end{document}